\documentclass[aps, prb, reprint, showpacs, twocolumn, 10pt, superscriptaddress, accepted=2018-05-21]{quantumarticle}

\usepackage{graphicx, amsmath, dsfont, dcolumn, bm, times, color, braket, ulem, bbold}

\usepackage[usenames,dvipsnames]{xcolor}

\usepackage[colorlinks]{hyperref}
\hypersetup{colorlinks, citecolor=blue, filecolor=blue, linkcolor=blue, urlcolor=blue}

\begin{document}

\bibliographystyle{apsrev4-1}

\title{Phonon-assisted relaxation and decoherence of singlet-triplet qubits in Si/SiGe quantum dots }

\author{Viktoriia Kornich}
\affiliation{Department of Physics, University of Basel, Klingelbergstrasse 82, CH-4056 Basel, Switzerland}
\affiliation{Department of Physics, University of Wisconsin--Madison, Madison, Wisconsin 53706, USA}

\author{Christoph Kloeffel}
\affiliation{Department of Physics, University of Basel, Klingelbergstrasse 82, CH-4056 Basel, Switzerland}

\author{Daniel Loss}
\affiliation{Department of Physics, University of Basel, Klingelbergstrasse 82, CH-4056 Basel, Switzerland}
\affiliation{CEMS, RIKEN, Wako, Saitama 351-0198, Japan}

\date{\today}

\begin{abstract}
We study theoretically the phonon-induced relaxation and decoherence of spin states of two electrons in a lateral double quantum dot in a SiGe/Si/SiGe heterostructure. We consider two types of singlet-triplet spin qubits and calculate their relaxation and decoherence times, in particular as a function of level hybridization, temperature, magnetic field, spin orbit interaction, and detuning between the quantum dots, using Bloch-Redfield theory. We  show that the magnetic field gradient, which is usually applied to operate the spin qubit, may reduce the relaxation time by more than an order of magnitude. Using this insight, we  identify an optimal regime where the magnetic field gradient does not affect the relaxation time significantly, and we propose regimes of longest decay times. We take into account the effects of one-phonon and two-phonon processes and suggest how our theory can be tested experimentally. The spin lifetimes we find here for Si-based quantum dots are significantly longer than the ones reported for their GaAs counterparts.
\end{abstract}

\maketitle

\let\oldvec\vec
\renewcommand{\vec}[1]{\ensuremath{\boldsymbol{#1}}}
\newcommand{\lsim}{\raisebox{-0.13cm}{~\shortstack{$<$ \\[-0.07cm]
      $\sim$}}~}
\section{Introduction}
\label{sec:Introduction}

Quantum dots (QDs) populated by electrons or holes are considered to be promising platforms for the physical realization of qubits for quantum computation \cite{loss:pra98,hanson:rmp07,kloeffel:arcmp13}. Much progress both in theory and experiment was made in studying GaAs-based QDs \cite{johnson:nature05, levy:prl02, taylor:natphys05, petta:science05, foletti:09, khaetskii:prl02, merkulov:prb02, shulman:science12, coish:prb04, bluhm:nature11, burkard:prb99, klinovaja:prb12,doherty:15, yoneda:prl14, scarlino:prl14, chesi:prb14, cerfontaine:prl14, biesinger:arxiv15, martins:arxiv15}.
However, recently Si- or Ge-based QDs attracted much attention. The reason is that in isotopically purified $^{28}$Si or isotopes of Ge with nuclear spin $0$ (e.g.\ $^{76}$Ge) decoherence sources characteristic to GaAs are absent, namely hyperfine interaction and spin orbit interaction (SOI) due to lattice-inversion asymmetry. Known schemes for spin qubits in Si or Ge are based on, e.g., hole spins in CMOS devices \cite{maurand:natcom16}, hut wires \cite{watzinger:nanolett16}, and Ge-Si core-shell nanowires \cite{kloeffel:prb11, hu:nnano12, maier:prb13, kloeffel:prb13, brauns:prb16}, donor electron spins \cite{laucht:apl14, tyryshkin:nature12, hile:apl15, watson:prl15, wang:arxiv17}, host \cite{pla:prl14} and donor \cite{steger:science12, saeedi:science13, usman:prb15, kane:nature98, hill:sa15, muhonen:arxiv17} nuclear spins, nuclear-electron spin qubits (Si:Bi) \cite{morley:natmater13}, qubits based on Si/SiO$_2$ structures \cite{muhonen:jpcm15, veldhorst:nature15, calderon:prl06}, and lateral QDs within the two-dimensional electron gas (2DEG) in Si/SiGe heterostructures \cite{takeda:apl13, kawakami:nnano14, zajac:arxiv15, takeda:sciadv16, zajac:science18, watson:nat18}. The sixfold degeneracy of conduction band valleys in Si can be an additional source of decoherence \cite{gamble:prb12, tahan:prb14} compared to GaAs. However, four of the six valleys get split off by a large energy of the order of a hundred meV in SiGe/Si/SiGe quantum wells because of the strain \cite{tahan:prb14, schaeffler:sst97, friesen:prb07, zwanenburg:rmp13}. Due to confinement, which may also be varied via electric fields, the twofold degeneracy of the remaining valleys is lifted, and reported valley splittings are of the order of 0.1--1\mbox{ meV} \cite{friesen:prb07, goswami:nature07, xiao:apl10, yang:natcom13, zwanenburg:rmp13}. For instance, electric control over the valley splitting for QDs in Si/SiO$_2$ was reported, and the presented energy range for the valley splitting is 0.3--0.8\mbox{ meV} \cite{yang:natcom13}. Therefore, it is possible to suppress the effect of many valleys in Si if the energies characteristic for the qubit subspace are small enough. 

Following the development in theory and experiment investigating the behavior of electron spin states in single and double quantum dots in Si \cite{prada:prb08, raith:prb12, yang:natcom13, maune:nature12, prance:prl12, culcer:prb09, hung:prb13, assali:prb11, culcer:prl12, simmons:prl11, tyryshkin:physicae06, tahan:prb02, glavin:prb03, wang:jap11, wang:prb10, zwanenburg:rmp13, veldhorst:nature15, kawakami:nnano14, gamble:prb12, tahan:prb14, xiao:apl10, takeda:apl13, zajac:arxiv15, takeda:sciadv16, zajac:science18, watson:nat18, goswami:nature07}, we study a lateral double quantum dot (DQD) which is formed in a Si/SiGe heterostructure and occupied by two electrons. We consider the relaxation and decoherence of the two-electron spin states due to phonons. Given the recent high interest in spin qubits at the $S$-$T_-$ anticrossing \cite{chesi:prb14, wong:prb15}, where $S$ is a spin singlet and $T_-$ a spin triplet with magnetic quantum number $m=-1$ for the spin component along the quantization axis, we investigate how the relaxation time $T_1$ and the decoherence time $T_2$ of such qubits depend on temperature for different kinds of hybridization of the singlet. We derive and analyze the dependence of $T_1$ and $T_2$ on the magnetic field gradient, which is usually applied in order to operate the spin qubits \cite{chesi:prb14,wu:pnas14, kawakami:nnano14, shin:prl10}. We further study the effects of one-phonon and two-phonon processes and suggest regimes where our theory can be tested experimentally. We also consider the $S$-$T_0$ spin qubit \cite{levy:prl02, taylor:natphys05, petta:science05} in two regimes: large detuning and small detuning, as it was done in our previous work on DQDs in GaAs/AlGaAs \cite{kornich:prb14}. Here, $T_0$ is the spin triplet with $m=0$. We investigate the dependence of $T_1$ and $T_2$ on temperature and on different system parameters which were not considered before.  

The paper is organized as follows. In Sec.~\ref{sec:Model}, we present the Hamiltonian of our model and a short description of the Bloch-Redfield theory. In Sec.~\ref{sec:STMinusQubit}, we study the relaxation and decoherence of $S$-$T_-$-based spin qubits. The case of $S$-$T_0$ spin qubits is discussed in Sec.~\ref{sec:ST0Qubit}. Additional decay channels for the studied qubits are listed in Sec.~\ref{sec:Comparison} and our conclusions follow in Sec.~\ref{sec:Conclusions}.

\section{Model}
\label{sec:Model}

\subsection{Hamiltonian}
\label{subsec:Hamiltonian}

We consider lateral DQDs in a Si/SiGe heterostructure grown along the crystallographic direction [001], which we also denote as the $z$ direction. The confinement in the plane perpendicular to $z$ is generated by the gates. The homogeneous magnetic field $\bm{B}$ is in this plane. An applied magnetic field gradient, which is usually produced via a micromagnet, enables control over the Bloch sphere of the spin qubit even in the absence of hyperfine and spin orbit interactions \cite{chesi:prb14, wu:pnas14, kawakami:nnano14, shin:prl10}.

The Hamiltonian of the system reads
\begin{eqnarray}
\tilde{H} &=& \sum_{j=1,2} \left( H_0^{(j)} + H_Z^{(j)} + \tilde{H}_{SOI}^{(j)} + H_{b}^{(j)} + H_{el-ph}^{(j)}\right) \nonumber \\ & & + H_C + H_{ph},
\label{eq:HamiltonianStart}
\end{eqnarray}
where $j$ labels the electrons, $H_0$ comprises the kinetic and potential energy of an electron in a DQD potential, $H_Z$ is the Zeeman term due to the external magnetic field, $\tilde{H}_{SOI}$ is the spin orbit interaction after a suitable transformation that accounts for the effect of higher-energy states \cite{khaetskii:prb00, aleiner:prl01, golovach:prl04, stano:prb05, stano:prl06}, $H_{b}$ is the term that describes the effect of the applied magnetic field gradient, $H_{el-ph}$ is the electron-phonon interaction, $H_C$ is the Coulomb repulsion, and $H_{ph}$ is the Hamiltonian of the phonon bath. The details and definitions of Eq.~\eqref{eq:HamiltonianStart} are presented in Ref.~\onlinecite{kornich:prb14} (Sec.~II and Appendix B), except for the applied magnetic field gradient and the electron-phonon interaction Hamiltonian which we provide below in general form for one electron. 

The applied magnetic field gradient acts on electrons similarly to the stabilized nuclear polarization in GaAs DQDs that produces a different Overhauser field for each QD. The Hamiltonian is therefore of the same form as for the hyperfine interaction and reads
\begin{equation}
H_{b}=\frac{\bm{b}\cdot\bm{\sigma}}{4}(\mathcal{P}_L-\mathcal{P}_R),
\end{equation}
where $\bm{b}$ appears due to the magnetic field gradient between the QDs, $\bm{\sigma}$ is the vector of Pauli matrices for the electron spin, and $\mathcal{P}_L$ and $\mathcal{P}_R$ are projectors for the left and right QD, respectively \cite{kornich:prb14}.

Now we consider the electron-phonon interaction in Si. In stark contrast to GaAs, only the deformation potential electron-phonon interaction is present in Si. Another important difference is that the conduction band minimum in bulk Si is sixfold degenerate. However, because of the strain in SiGe/Si/SiGe quantum wells this sixfold degeneracy is lifted and there are only two degenerate valleys of lowest energy \cite{zwanenburg:rmp13, tahan:prb14, schaeffler:sst97, friesen:prb07}. These two minima in the conduction band are found at the wave vectors $\bm{k}_{+z} = k_0 \bm{e}_z$ and $\bm{k}_{-z} = - k_0 \bm{e}_z$, where $\bm{e}_z$ is the unit vector along the $z$ direction, i.e., the [001] direction, and $k_0 \simeq 9.5\mbox{ nm$^{-1}$}$ \cite{friesen:prb07}. The confinement, tunable by electric fields, lifts the last degeneracy, and in good approximation the two $z$~valleys at $\bm{k}_{\pm z}$ are the only valleys involved in the low-energy electron states \cite{zwanenburg:rmp13, goswami:nature07, tahan:prb14, friesen:prb07}. Therefore, following Refs.~\onlinecite{herring:pr56, yu:book10}, the electron-phonon Hamiltonian for our system reads
\begin{equation}
H_{el-ph}=\Xi_d\, \mbox{Tr}\, \bm{\varepsilon} +\Xi_u\, \bm{e}_z\cdot \bm{\varepsilon} \cdot \bm{e}_z .
\label{eq:HelphLongForm}
\end{equation}
Here, $\bm{\varepsilon}$ is a strain tensor defined as
\begin{equation}
\varepsilon_{ij}=\frac{1}{2}\left(\frac{\partial u_i}{\partial r_j}+\frac{\partial u_j}{\partial r_i}\right),
\end{equation}
where $i,j$ denote the spatial components, $\bm{r}$ is the position in the material, and $\bm{u}$ is the displacement operator. The trace of the strain tensor is $\mbox{Tr}\, \bm{\varepsilon} $.
We note that the electron-phonon Hamiltonian of Eq.~\eqref{eq:HelphLongForm} is equivalent to
\begin{equation}
H_{el-ph}=\Xi_d\mbox{Tr}\, \bm{\varepsilon} + \Xi_u\varepsilon_{zz}.
\label{eq:H_elph_Si_finalform}
\end{equation}
The displacement operator can be represented in the form \cite{kornich:prb14}
\begin{equation}
\bm{u} = \sum_{\bm{q},s} \sqrt{\frac{\hbar}{2 \rho V q v_s}} \bm{e}_{\bm{q}s} \left( a_{\bm{q}s} \mp_s a_{-\bm{q}s}^\dagger \right) e^{i \bm{q} \cdot \bm{r}},
\end{equation}
where $s \in \{l, t_1, t_2\}$ stands for the longitudinal and transverse acoustic modes, $\bm{q}$ is a wave vector within the first Brillouin zone, and $q = |\bm{q}|$. We choose the normalized polarization vectors such that $\bm{e}_{\bm{q}l} = \bm{q}/q$, $\bm{e}_{-\bm{q}t_1} = -\bm{e}_{\bm{q}t_1}$, $\bm{e}_{-\bm{q}t_2} = \bm{e}_{\bm{q}t_2}$, and so $\mp_l = - =\mp_{t_1}$ and $\mp_{t_2} = +$. A phonon with properties $\bm{q}$ and $s$ is annihilated and created by the operators $a_{\bm{q}s}$ and $a_{\bm{q}s}^\dagger$, respectively. We note that the transverse phonons do not contribute to the term $\Xi_d\mbox{Tr}\, \bm{\varepsilon}$ in Eq.~\eqref{eq:H_elph_Si_finalform}, they only contribute to $\Xi_u\varepsilon_{zz}$. For further calculations we use the density $\rho = 2.33\mbox{ g/cm$^3$}$, the deformation potential constants \cite{yu:book10} $\Xi_d=5\mbox{ eV}$ and $\Xi_u=8.77\mbox{ eV}$, and the averaged sound velocities \cite{cleland:book, adachi:properties} $v_l=9\times 10^3\mbox{ m/s}$, $v_{t_1}=v_{t_2}=5.4\times 10^3\mbox{ m/s} = v_t$. Later on, when we calculate the qubit lifetimes, we can use the continuum limit and we integrate to infinite $q$ for convenience, as terms with $\bm{q}$ outside the first Brillouin zone do not affect our results for the temperatures considered here. The sample volume $V$ will cancel out in the analysis.

We wish to emphasize that the results presented in this work cannot be obtained by simply replacing the parameters of our previous calculations \cite{kornich:prb14} for GaAs DQDs, even though several parts of the model can be adopted. The reason is that the electron-phonon interaction in Si fundamentally differs from that in GaAs or similar materials. In fact, we find that the novel term $\Xi_u\varepsilon_{zz}$ in Eq.~\eqref{eq:H_elph_Si_finalform} is crucial for the phonon-mediated decay of singlet-triplet qubits in Si DQDs.

\subsection{Basis states and projected Hamiltonian}
\label{subsec:BasisStates}

In this subsection we consider the Hamiltonian [Eq.~\eqref{eq:HamiltonianStart}] in the basis $\{\ket{(1,1)T_0}$, $\ket{(1,1)S}$, $\ket{(1,1)T_+}$, $\ket{(1,1)T_-}$, $\ket{(0,2)S}$, $\ket{(2,0)S}\}$, where the first and second indices in parentheses correspond to the occupation number of the left and right QD, respectively, $S$ denotes spin singlet states, and $T$ denotes spin triplet states. As each minimum of the DQD potential in the plane of the 2DEG is well approximated by a 2D harmonic oscillator potential, we use linear combinations of the harmonic oscillator eigenfunctions to describe the in-plane orbital part of the electron state in the DQD potential \cite{burkard:prb99}. For the explicit expressions of wave functions and details see Appendix~A of Ref.~\onlinecite{kornich:prb14}. Due to the strong confinement of the electrons in the growth direction, it turns out that the wave functions chosen along $z$ hardly affect the phonon-assisted relaxation and decoherence processes that we are interested in. In contrast to Ref.~\onlinecite{kornich:prb14}, where a triangular potential based on typical GaAs/AlGaAs heterostructures was assumed, we consider a SiGe/Si/SiGe quantum well and approximate it by a hard-wall potential 
\begin{equation}
V(z) = \left\{ \begin{array}{ll}
\infty, & z < 0, \\
C , & 0<z<a_z,\\
\infty, & z > a_z,
\end{array} \right.
\end{equation}
\\
where $C$ is a constant with units of energy and $z = 0,\ a_z$ corresponds to the interface between SiGe ($z<0, \ z>a_z$) and Si ($0<z<a_z$). The ground state wave function in such a potential is
\begin{equation}
\phi_z(z) = \sqrt{\frac{2}{a_z}}\sin{\left[\frac{\pi z}{a_z}\right]} ,
\label{eq:FangHoward}
\end{equation}
with $a_z$ being a positive length that is interpreted as the width of the 2DEG in $z$ direction. We take $a_z=6\mbox{ nm}$ for all numerical calculations in this work. We note that in experiments an electric field is usually applied along the growth direction of Si/SiGe heterostructures, which changes the shape of the assumed quantum well potential from rectangular toward triangular. However, as the electrons are strongly confined along~$z$, the details of the well hardly affect the qubit lifetimes and we find that our results do not change by more than $\sim$10\% when the potential becomes completely triangular. As a consequence, our theory is also well applicable to, e.g., lateral Si DQDs formed in Si/SiO$_2$ systems. 

The Hamiltonian in our basis reads
\begin{widetext}{\small
\begin{eqnarray}
\label{eq:HamiltonianSmall}
\tilde{H}&=&\begin{pmatrix}P_T & \frac{b_B}{2} & 0 & 0 & 0 & 0 \\
\frac{b_B}{2} & V_+-V_-+P_T & \frac{\Omega}{\sqrt{2}}-\frac{1}{2\sqrt{2}}(b_{x}+i b_{z}) & -\frac{\Omega}{\sqrt{2}}+\frac{1}{2\sqrt{2}}(b_{x}-i b_{z}) & -\sqrt{2}t+P_S^\dagger & -\sqrt{2}t+P_S\\
0 & \frac{\Omega}{\sqrt{2}}-\frac{1}{2\sqrt{2}}(b_{x}-i b_{z}) & E_Z+P_T & 0 & 0 & 0\\
0 & -\frac{\Omega}{\sqrt{2}}+\frac{1}{2\sqrt{2}}(b_{x}+i b_{z}) & 0 & -E_Z+P_T & 0 & 0 \\
0 & -\sqrt{2}t+P_S & 0 & 0 & -\epsilon+U-V_-+P_{SR} & 0\\
0 & -\sqrt{2}t+P_S^\dagger & 0 & 0 & 0 & \epsilon+U-V_-+P_{SL}\\ \end{pmatrix} \nonumber \\ 
& & + H_{ph},
\end{eqnarray}}
\hspace*{-0.09cm}where $t$ is the tunnel coupling, $U$ is the onsite repulsion, $V_+$ and $V_-$ are the matrix elements of Coulomb interaction, and $E_Z=g \mu_B B$ is the Zeeman splitting with $B=|\bm{B}|$, the Bohr magneton $\mu_B$, and $g=2$ as the electron $g$-factor in Si. The terms $b_{x}$, $b_{z}$, and $b_B$ are produced by an applied magnetic field gradient along the $x$, $z$ axes and $\bm{B}$, respectively, where the three orthogonal directions for $x$, $z$, and $\bm{B}$ form a right-handed basis (meaning that $\bm{B}$ points in the negative $y$ direction, given that the axes $x$, $y$, $z$ belong to a right-handed coordinate system). For simplicity, we set $b_z=0$ in the following, as it can be achieved experimentally using a micromagnet. The electrical bias (detuning) between the dots is denoted by $\epsilon$, where $\epsilon=0$ is for the unbiased DQD \cite{stepanenko:prb12}.

The electron-phonon matrix elements are
\begin{align}
P_T =& \bra{(1,1)S}\sum_{j=1,2}H_{el-ph}^{(j)}\ket{(1,1)S}=\bra{(1,1)T_0}\sum_{j=1,2}H_{el-ph}^{(j)}\ket{(1,1)T_0},\\ 
P_S =& \bra{(1,1)S}\sum_{j=1,2}H_{el-ph}^{(j)}\ket{(2,0)S}=\bra{(0,2)S}\sum_{j=1,2}H_{el-ph}^{(j)}\ket{(1,1)S}, \\
P_{SR} =& \bra{(0,2)S}\sum_{j=1,2}H_{el-ph}^{(j)}\ket{(0,2)S},\\
P_{SL} =& \bra{(2,0)S}\sum_{j=1,2}H_{el-ph}^{(j)}\ket{(2,0)S}.
\end{align}
\end{widetext}
As evident from the provided equations, these matrix elements have a similar structure but differ due to the integrals for the orbital parts. We note that the matrix elements all commute with each other, even though they still contain the creation and annihilation operators for the phonons.   

The matrix element $\Omega$ comes from SOI and has the form \cite{stano:prl06, kornich:prb14}
\begin{equation}
\Omega=\frac{F(L,l_c)g\mu_B B}{l_R}\cos{\eta}.
\label{eq:SOI}
\end{equation}
The function $F(L,l_c)$ depends on the distance $L$ between the centers of the QDs and the confinement length $l_c=\sqrt{\hbar^2/(m_{\rm eff}\Delta E)}$, where $m_{\rm eff}=1.73 \times 10^{-31}\mbox{ kg}$ is the transverse effective mass of an electron in Si and $\Delta E$ is the orbital level spacing in each QD. We note that $l_c$ determines the Gaussian decay of the wave functions and $F(L,l_c) \approx -L$ when the dots are only weakly coupled. The Rashba length $l_R = \hbar/(m_{\rm eff}\alpha)$ is related to the SOI amplitude $\alpha$ in the Rashba Hamiltonian, and $\eta$ is the angle between $\bm{B}$ and the axis that connects the two QDs.

\subsection{Bloch-Redfield theory}
\label{subsec:BlochRedfieldTheory}

To calculate the relaxation time $T_1$ and decoherence time $T_2$ we use the Bloch-Redfield theory \cite{golovach:prl04, slichter:book, borhani:prb06}, which describes the dynamics of the qubit interacting with the bath of phonons. The Bloch-Redfield formalism makes use of a Markov approximation, which is very well justified for our system because the qubit lifetimes $T_1$ and $T_2$ are several orders of magnitude longer than the correlation time~$\tau_c$. The latter can be estimated by considering the maximal duration which a phonon needs to travel through the DQD \cite{golovach:prl04}. With the parameters used in the present work, this estimate yields $\tau_c \lsim (L + 2 l_c)/v_{t} \approx 0.04 \mbox{ ns}$.

In the following we will consider $S$-$T_-$ and $S$-$T_0$ qubits. To decouple the qubit subspace from the other states, we first apply a unitary transformation to $\tilde{H}$ that diagonalizes $\tilde{H}-H_{el-ph}^{(1)}-H_{el-ph}^{(2)}$, where $(1)$ and $(2)$ label the first and second electron, respectively. The transformation matrix for this first step is found numerically. Depending on the qubit under study ($S$-$T_0$ or $S$-$T_-$), we then perform a Schrieffer-Wolff transformation up to the third order to take into account both one-phonon and two-phonon processes. The Schrieffer-Wolff transformation corresponds to quasi-degenerate perturbation theory \cite{winkler:book}. It is valid when the matrix elements resulting from electron-phonon coupling are much smaller than the energy difference between the qubit subspace and other states. After the Schrieffer-Wolff transformation, the qubit subspace is well separated from other states, which allows us to study the dynamics in terms of an effective Hamiltonian of the form $H_q + H_{q-ph}(\tau) + H_{ph}$, where $H_q$ is a 2$\times$2 part that contains information about the qubit without phonons and $H_{q-ph}(\tau)$ is the interaction between qubit and phonons at time $\tau$ in the interaction representation. 

Defining the pseudo-spin vector $\bm{\tilde{\sigma}}$ as a vector of Pauli matrices $\sigma_{\tilde{x}}, \sigma_{\tilde{y}}, \sigma_{\tilde{z}}$ in the qubit subspace, where $\tilde{x}$, $\tilde{y}$, $\tilde{z}$ are the directions in the pseudo-spin space, we can represent $H_q$ and $H_{q-ph}(\tau)$ as
\begin{gather}
H_q = B_{\rm eff} \sigma_{\tilde{z}},\\
H_{q-ph}(\tau) = \delta\bm{B}(\tau)\cdot \bm{\tilde{\sigma}} ,
\end{gather}
where $B_{\rm eff}$ is a positive energy and $\delta \bm{B}(\tau)$ contains electron-phonon interaction matrix elements. The expressions for $B_{\rm eff}$ and $\delta \bm{B}(\tau)$ result from all linear transformations performed before and are newly calculated whenever the input parameters or the qubit type change. Following the theory from Refs.~\onlinecite{golovach:prl04, borhani:prb06}, the times $T_2$ and $T_1$ are 
\begin{gather}
\label{eq:T2}
\frac{1}{T_2} = \frac{1}{2 T_1} + \frac{1}{T_\varphi}, \\ 
\label{eq:T1}
\frac{1}{T_1} = J_{\tilde{x}\tilde{x}}^+(\Delta_{ST})+J_{\tilde{y}\tilde{y}}^+(\Delta_{ST}) , \\ 
\label{eq:Tphi}
\frac{1}{T_{\varphi}} = J_{\tilde{z}\tilde{z}}^+(0) .
\end{gather} 
The quantity $\Delta_{ST}$ is defined as $\Delta_{ST}= 2 B_{\rm eff}$, i.e., the energy splitting between qubit states without taking into account electron-phonon interaction, and 
\begin{equation}
\label{eq:Jplus}
J_{ii}^+(\hbar\omega)=\frac{2}{ \hbar^2}\int_{-\infty}^{\infty}\cos(\omega \tau)\langle\delta B_i(0)\delta B_i (\tau)\rangle d\tau .
\end{equation}
The correlator $\langle \delta B_i(0)\delta B_i (\tau) \rangle$ is evaluated for a phonon bath in thermal equilibrium at temperature $T$. The time $T_\varphi$ represents the pure dephasing part in the decoherence time $T_2$.

\subsection{Corrections from valley degrees of freedom}
\label{subsec:CorrectionsFromValleys}

When analyzing electrons in Si, it is important to consider possible effects of the valley degrees of freedom. In particular, the presence of more than one valley can have notable effects when excited states, which differ from the ground state by more than just the spin part, are occupied. For instance, this applies to singlet-triplet qubits in a single QD. When the qubit is in the triplet state, the two electron spins are parallel. Consequently, because of the Pauli exclusion principle, one of the two electrons must occupy an excited state. Compared with the ground state of the QD, the occupied excited state may differ in the valley or in the orbital part. In such systems, it therefore matters whether the orbital level spacing is smaller or greater than the valley splitting \cite{gamble:prb12, yang:natcom13}.

In the present work we study singlet-triplet qubits in a DQD instead of a single QD. Compared with the latter, the choice between the left and the right QD of the DQD provides an additional degree of freedom. Therefore, in the limit where the two QDs have zero overlap, two electrons in the DQD can be in spin singlet and triplet states by always occupying the ground states of the QDs as far as the valley and orbital degrees of freedom are concerned. Our work focuses on singlet-triplet qubits which are based on the low-energy eigenstates of two electrons in a DQD with weakly coupled QDs ($L > 2 l_c$), such that the Hund-Mulliken approach is applicable \cite{burkard:prb99, stepanenko:prb12}. Consequently, it is not important in our model whether excited states in the QDs feature an excitation of orbital or valley type. There is only one exception where we will make the assumption that the valley splitting is larger than the orbital level spacing, see Sec.~\ref{subsec:SmallDetuning}. While also the opposite case is possible \cite{yang:natcom13, gamble:prb12}, we note that the assumption we will make in Sec.~\ref{subsec:SmallDetuning} was made in earlier calculations \cite{gamble:prb12, prada:prb08} and that the realization of this regime has already been reported \cite{xiao:apl10}. 

When studying the qubit decay in coupled Si QDs, using basis states without a valley degree of freedom (as in GaAs) is usually considered as a reasonable approximation \cite{srinivasa:prb16, hu:prb11}. Provided that the valley splitting is large, which is feasible given the recent experimental progress \cite{friesen:prb07, goswami:nature07, xiao:apl10, yang:natcom13, zwanenburg:rmp13}, we therefore omit the valley degree of freedom in our calculations. If details about the valleys and the splittings are known, our model may be extended by choosing a larger set of basis states that takes the valleys into account \cite{rohling:njp12}. As long as the valley splitting corresponds to a large quantity, however, we believe that inclusion of the valley states cannot lead to qubit lifetimes that are significantly shorter than those presented here. For corrections from additional decay mechanisms, we refer to Sec.~\ref{sec:Comparison}.

\section{$S$-$T_-$ qubit}    
\label{sec:STMinusQubit}

Experimental and theoretical studies with two electrons in GaAs DQDs have shown that the $S$-$T_+$ anticrossing is useful for quantum information processing \cite{ribeiro:prb10, petta:science10, ribeiro:prl13, ribeiro:prb13, chesi:prb14}. For instance, a scheme has been proposed where single-spin rotations are performed at the center of the $S$-$T_+$ anticrossing \cite{chesi:prb14}. One of the two electron spins serves as an ancillary spin in this scheme, and short gate times of a few nanoseconds were calculated for realistic setups with a micromagnet. Since the electron $g$-factor is negative in GaAs but positive in Si, such schemes which are based on the $S$-$T_+$ anticrossing in GaAs DQDs may easily be transferred to the $S$-$T_-$ anticrossing in Si DQDs \cite{wu:pnas14}. Moreover, high gate fidelities have recently been predicted for singlet-triplet qubits in Si DQDs operated at the center of the $S$-$T_-$ anticrossing \cite{wong:prb15}. An advantage of the anticrossing is that $\partial \Delta_{ST}/\partial \epsilon \simeq 0$ near its center, and so the qubit is to a certain extent protected against charge noise \cite{chesi:prb14, wong:prb15}, especially when the anticrossing region is relatively wide.

\begin{figure}[tb] 
\begin{center}
\includegraphics[width=\linewidth]{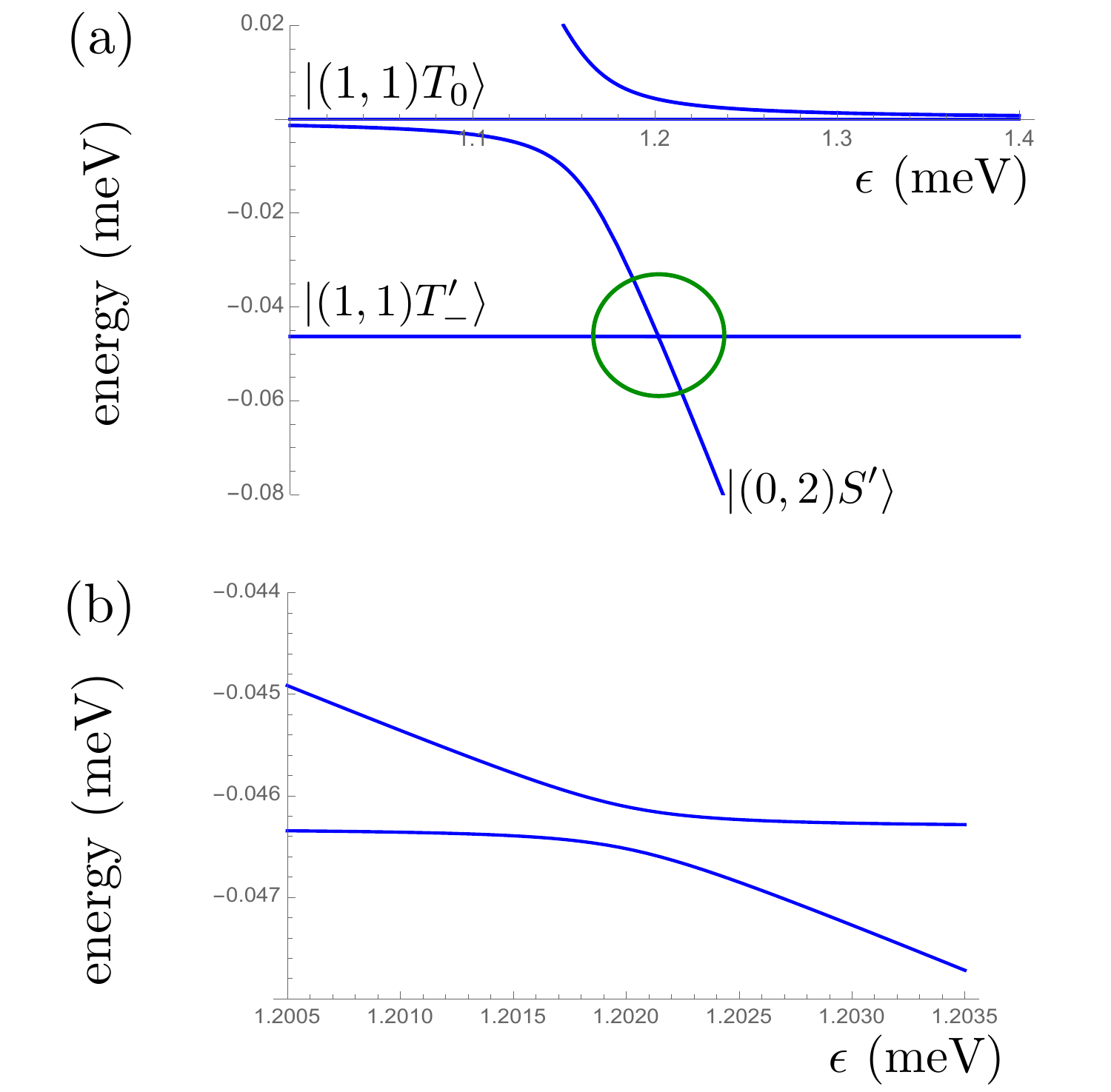}
\caption{
The energy spectrum of two-electron states in a double quantum dot as a function of detuning $\epsilon$ denoting the energy difference between the two dots. The green circle in panel a shows the $S$-$T_-$ anticrossing, which is shown enlarged in panel b. The parameters used are the same as for Fig.~\ref{fig:STminus_T}. At the anticrossing, the singlet state is of type $\ket{(0,2)S}$.} 
\label{fig:S_qubit_Si}
\end{center}
\end{figure}

\begin{figure}[tb] 
\begin{center}
\includegraphics[width=\linewidth]{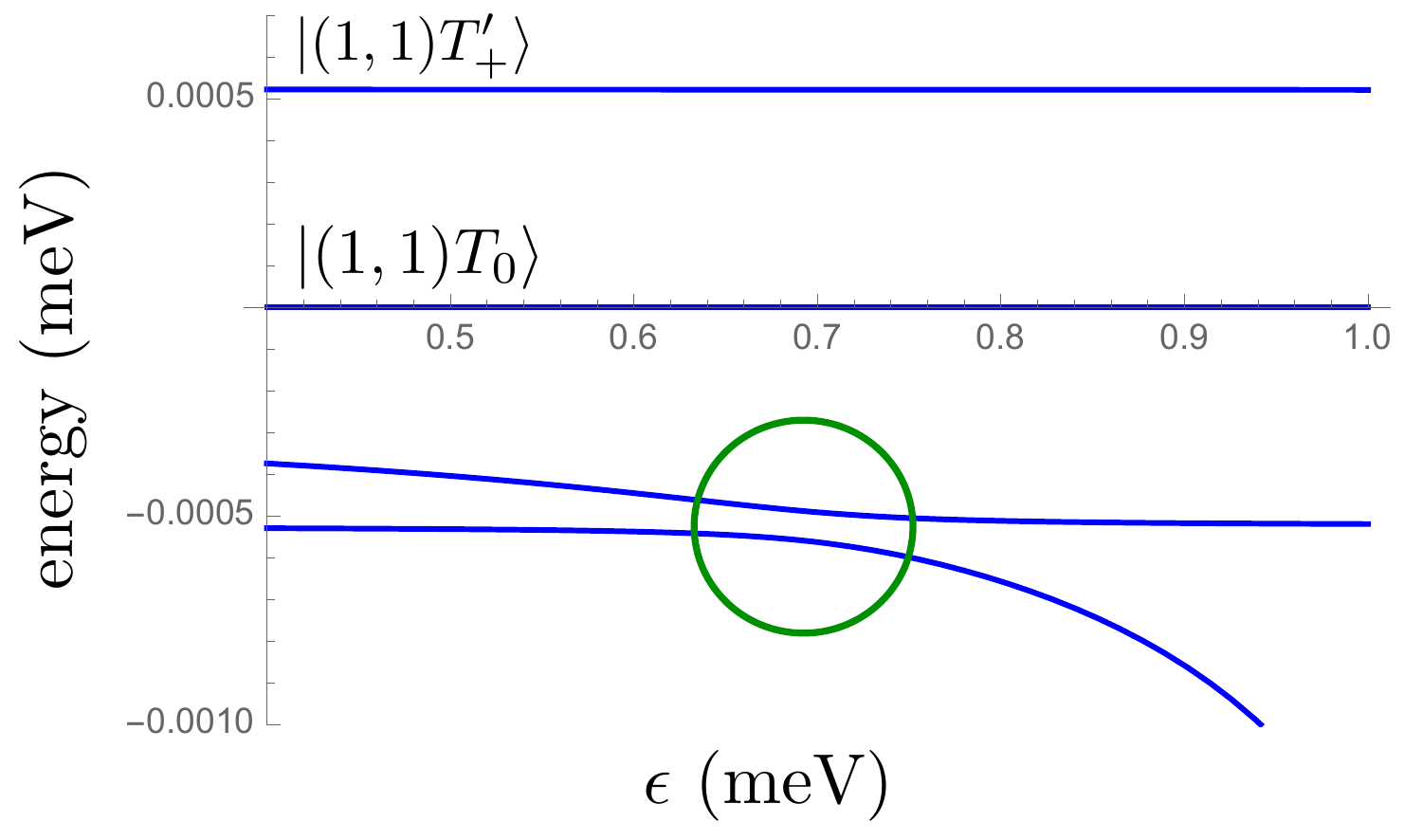}
\caption{The energy spectrum of two-electron states in a double quantum dot as a function of detuning $\epsilon$. The green circle shows the $S$-$T_-$ anticrossing. The parameters used are the same as for Fig.~\ref{fig:S11Tminus_T}. At the anticrossing, the singlet state is of type $\ket{(1,1)S}$. }
\label{fig:F_qubit_Si}
\end{center}
\end{figure}

Following the interest in building a qubit based on the $S$-$T_-$ anticrossing \cite{chesi:prb14, wong:prb15} we study phonon-induced relaxation and decoherence for the electron spin states at this anticrossing. Because of tunnel coupling, magnetic field gradient, and SOI all the states in our basis are hybridized to some extent. We therefore consider two possible regimes: the state that mainly consists of $\ket{(1,1)T_-}$ (we denote it as $\ket{(1,1)T_-'}$) anticrosses with the state that is mainly $\ket{(0,2)S}$ ($\ket{(0,2)S'}$) or mainly $\ket{(1,1)S}$ ($\ket{(1,1)S'}$). This can also be seen from the spectrum. In Fig.~\ref{fig:S_qubit_Si}a we plotted the dependence of the energy of two-electron states on the detuning $\epsilon$. The green circle highlights the region of anticrossing between $\ket{(1,1)T_-'}$ and $\ket{(0,2)S'}$, which is shown enlarged in Fig.~\ref{fig:S_qubit_Si}b. In Fig.~\ref{fig:F_qubit_Si} we choose different  parameters and show the anticrossing between $\ket{(1,1)T_-'}$ and $\ket{(1,1)S'}$. When plotting these spectra, the electron-phonon interaction was omitted.

\subsection{The qubit based on $\ket{(1,1)T_-'}$-$\ket{(0,2)S'}$} 
\label{subsec:Tminus02S}

Here we study the case shown in Fig.~\ref{fig:S_qubit_Si}.

\subsubsection{Dependence on temperature}
\label{subsubsec:Dep_T_Qubit11Tprime02Sprime}

\begin{figure}[tb] 
\begin{center}
\includegraphics[width=\linewidth]{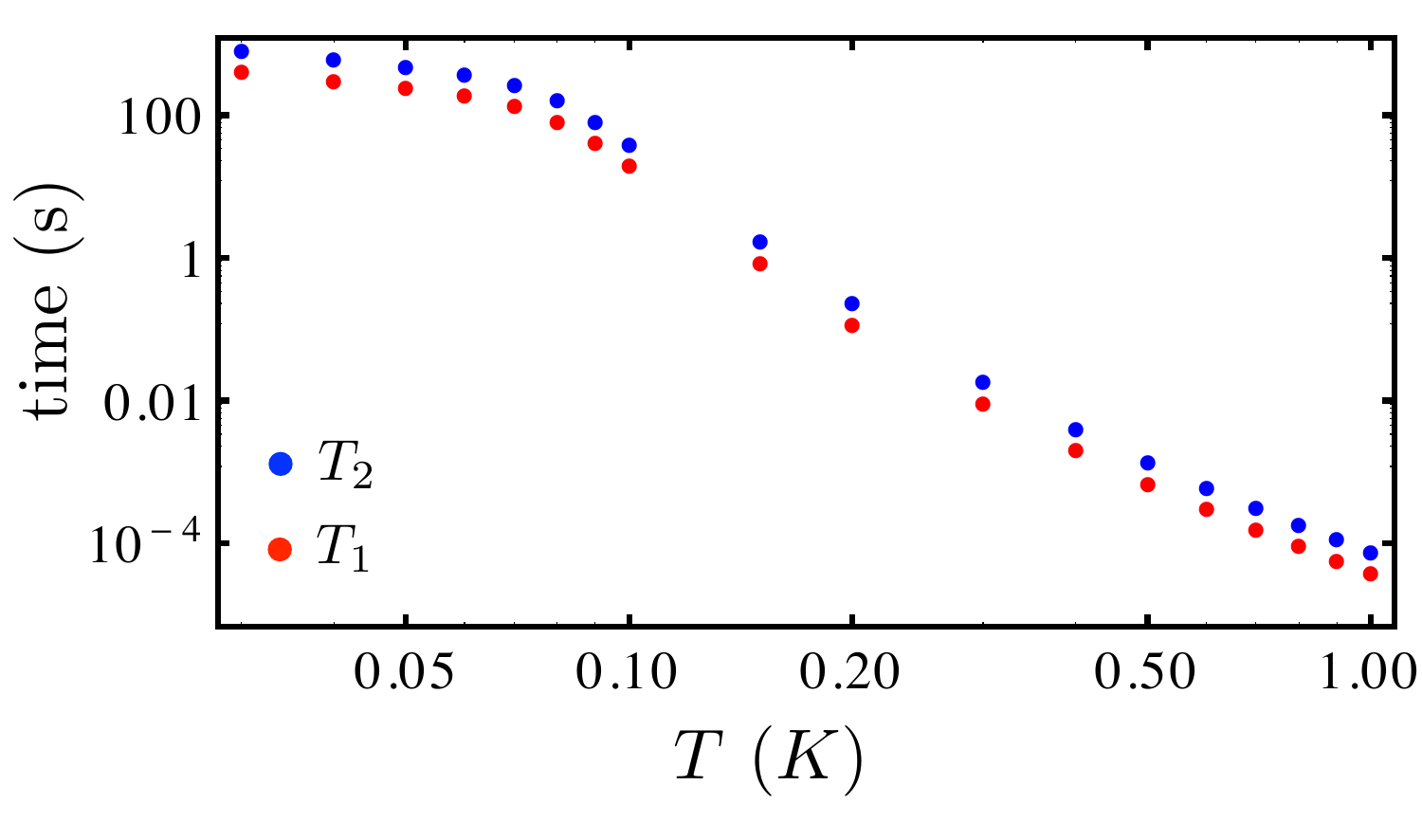}
\caption{The dependence of $T_1$ (red) and $T_2$ (blue) on temperature for the parameters listed in Sec.~\ref{subsubsec:Dep_T_Qubit11Tprime02Sprime}. The anticrossing is between $\ket{(1,1)T_-'}$ and $\ket{(0,2)S'}$.}
\label{fig:STminus_T}
\end{center}
\end{figure}

We plot the dependence of $T_1$ and $T_2$ on temperature~$T$ in Fig.~\ref{fig:STminus_T}, for which we used the following parameters: $B=0.4\mbox{ T}$, $t=10\mbox{ $\mu$eV}$, $V_+=40\mbox{ $\mu$eV}$, $V_-=39.99\mbox{ $\mu$eV}$, $U=1.2\mbox{ meV}$, $b_x=2\mbox{ $\mu$eV}$, $L=150\mbox{ nm}$, $l_c=42.7\mbox{ nm}$ (i.e., $\Delta E=200\mbox{ $\mu$eV}$), and $\epsilon=1.201988\mbox{ meV}$. The region where $\ket{(1,1)T_-'}$ and $\ket{(0,2)S'}$ anticross is typically quite narrow, and therefore such a high precision in $\epsilon$ is needed to operate exactly in the anticrossing center. That is the point where, if we take $b_x - 2 \Omega = 0$, the energies of $\ket{(0,2)S'}$ and $\ket{(1,1)T_-'}$ are equal, i.e., $\ket{(0,2)S'}$ and $\ket{(1,1)T_-}$ cross. We take $b_B=0$ to decouple the qubit subspace from $\ket{(1,1)T_0}$ \cite{wong:prb15}.

As SOI enters in Eq.~\eqref{eq:HamiltonianSmall} together with $b_x$, we neglect it assuming $|\Omega| \ll |b_x|$. In lateral SiGe/Si/SiGe QDs, SOI might be due to QD confinement or other applied electric fields, imperfections of the quantum well \cite{wilamowski:prb02}, or interface effects between two semiconductors \cite{vervoort:prb97, vervoort:sst99}. According to Ref.~\onlinecite{wilamowski:prb02}, the spin-orbit length is $l_R= 73\mbox{ $\mu$m}$. Using this value, we get $\Omega=-0.095 \mbox{ $\mu$eV}$ from Eq.~\eqref{eq:SOI} when $\eta=0$. The SOI due to interface effects between semiconductors is absent if the amount of atomic monolayers of Si is even \cite{nestoklon:prb08}. However, experimental values for any of the three origins mentioned above are not known to us for present-day samples.  

\begin{figure}[tb] 
\begin{center}
\includegraphics[width=\linewidth]{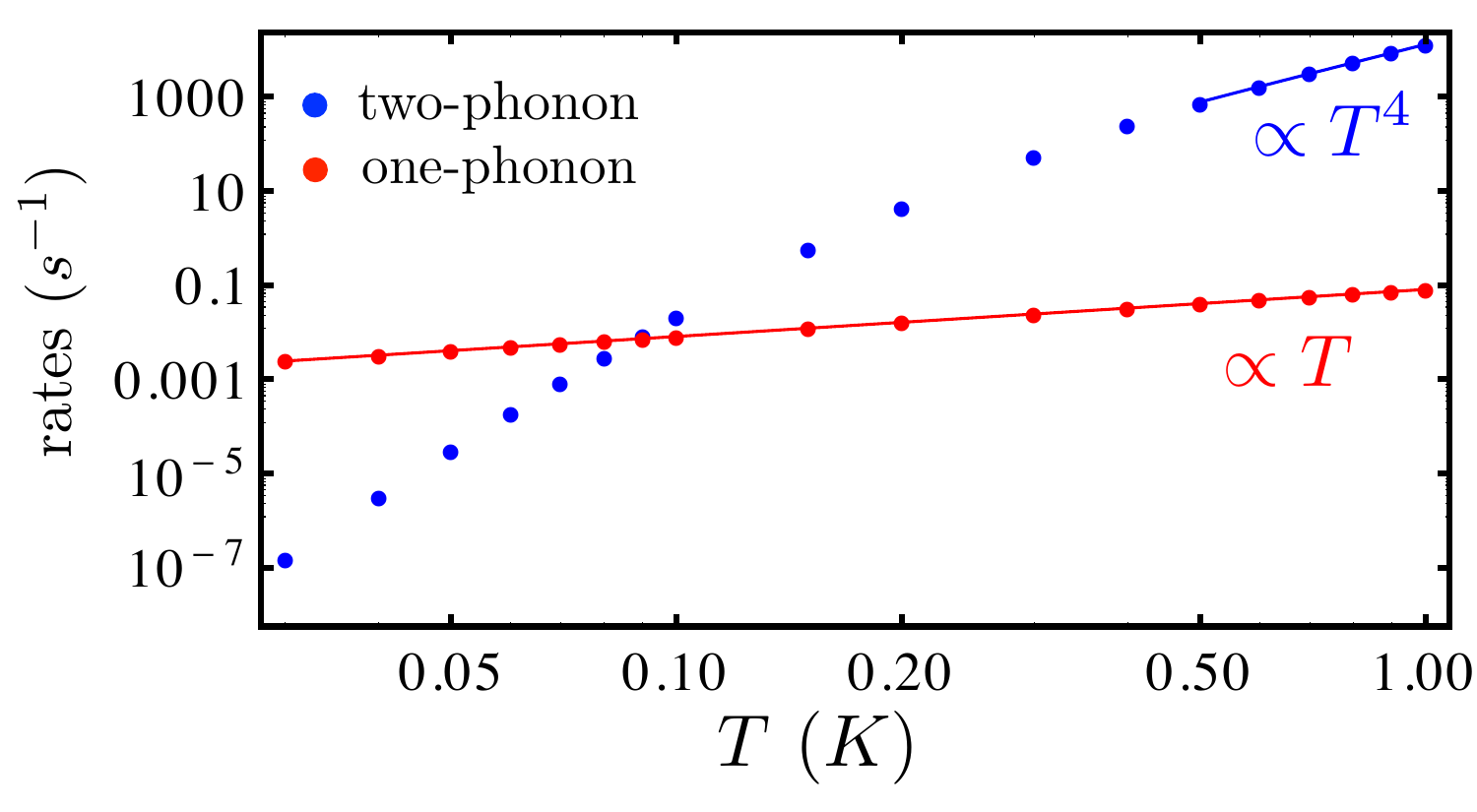}
\caption{The dependence of one-phonon ($\Gamma_2^{1p}$, red) and two-phonon ($\Gamma_2^{2p}$, blue) components of the decoherence rate $1/T_2 = \Gamma^{1p}_2 + \Gamma^{2p}_2$ on temperature. The straight lines obey the shown power-laws and are fits to the numerical data. The parameters are the same as in Fig.~\ref{fig:STminus_T}. }
\label{fig:STminus_rates_T}
\end{center}
\end{figure}

In Fig.~\ref{fig:STminus_T} we see that $T_2\simeq 2T_1$, that means the relaxation part dominates over dephasing in $T_2$ [see Eq.~\eqref{eq:T2}]. Up to a temperature of $T=0.08\mbox{ K}$, both $T_1$ and $T_2$ decay slowly and then change their behavior to a more rapid decay. To explain this change at around $T\simeq 0.08\mbox{ K}$ we plot the one-phonon process decoherence rate ($\Gamma^{1p}_2$) and two-phonon process decoherence rate ($\Gamma^{2p}_2$), see Fig.~\ref{fig:STminus_rates_T}. These rates contribute to $T_2$ as
\begin{equation}
\frac{1}{T_2} = \Gamma_2 =\Gamma^{1p}_2 + \Gamma^{2p}_2.
\end{equation}
We note that the one-phonon process can lead only to relaxation, it cannot lead to dephasing \cite{kornich:prb14}. 
Therefore $\Gamma_2^{1p} = \Gamma_1^{1p}/2$, where $\Gamma_1^{1p}$ is the one-phonon process relaxation rate. From Fig.~\ref{fig:STminus_rates_T} we can see that $\Gamma_2^{1p}\propto T$ for the whole temperature range. This is so because the dominant terms in $\Gamma_2^{1p}$ are proportional to a Bose-Einstein distribution,
\begin{equation}
\Gamma_2^{1p}\propto ({e^{\frac{\Delta_{ST}}{k_BT}}-1})^{-1}\simeq \frac{k_BT}{\Delta_{ST}}.
\label{eq:Gamma_2_1p}
\end{equation}
In our case $\Delta_{ST}\ll k_B T$ for all temperatures under consideration, therefore the second equality in Eq.~\eqref{eq:Gamma_2_1p} is justified. The two-phonon process rate $\Gamma_2^{2p}$ has a more complicated dependence on temperature. For $0.03\mbox{ K}<T<0.07\mbox{ K}$, we find $\Gamma_2^{2p} \simeq C_1 + C_2 T^9$, where $C_1$ and $C_2$ are constants, then $\Gamma_2^{2p}$ grows more slowly, and for $0.5\mbox{ K}<T<1\mbox{ K}$ we obtain $\Gamma_2^{2p}\propto T^4$ in good approximation. Consequently, the change in the decay of $T_1$ and $T_2$ at $T\simeq 0.08\mbox{ K}$ in Fig.~\ref{fig:STminus_T} is due to the fact that for lower temperatures the relaxation happens mainly via one-phonon processes, with rate $\propto T$, and for higher temperatures two-phonon processes dominate with the rate depending on higher powers of $T$. The crossover between these two regimes occurs at around $0.08\mbox{ K}$ (see Fig.~\ref{fig:STminus_rates_T}).

\subsubsection{Dependence on the magnetic field gradient}
\label{subsubsec:Dep_bB_Qubit11Tprime02Sprime}

\begin{figure}[tb] 
\begin{center}
\includegraphics[width=\linewidth]{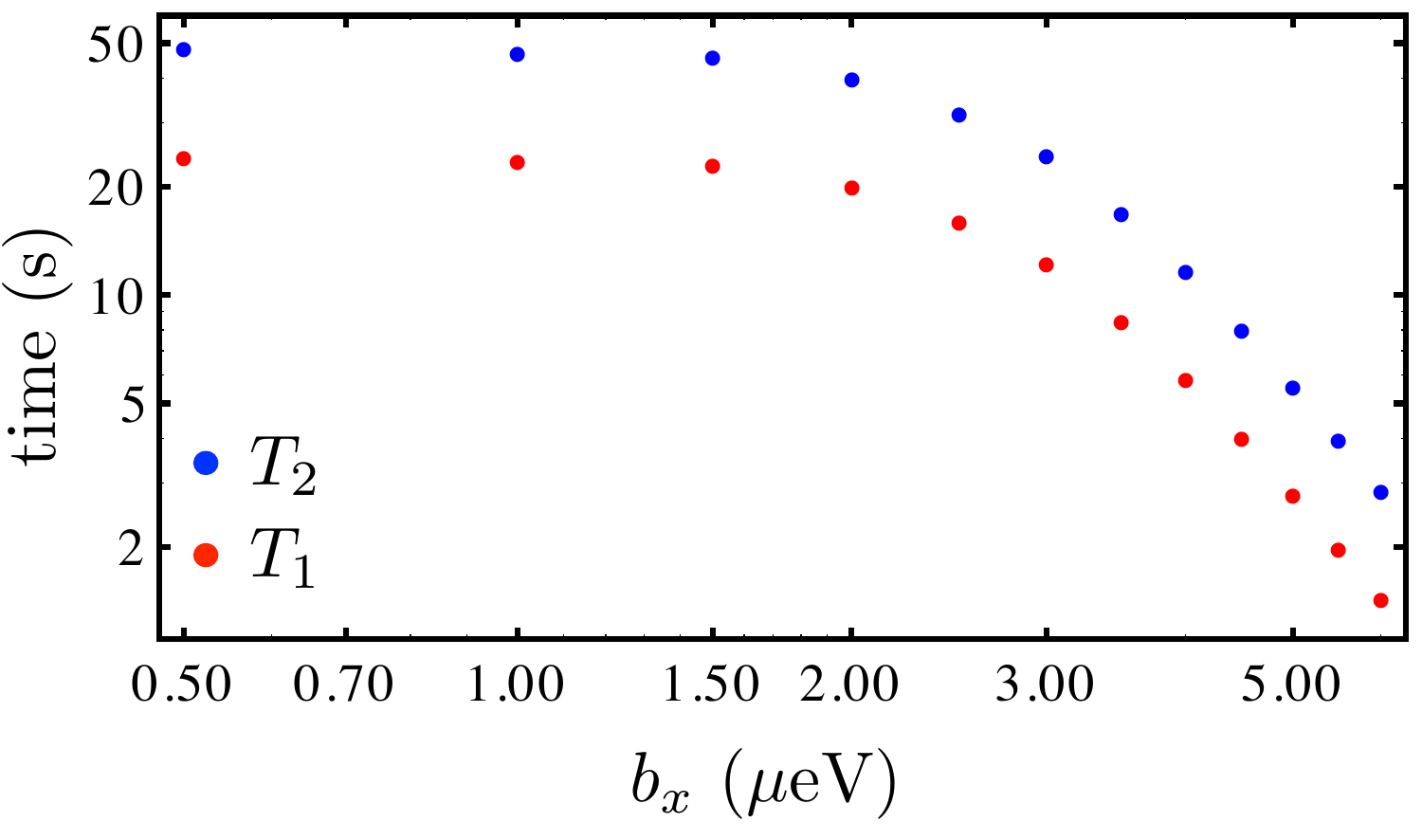}
\caption{The dependence of $T_1$ (red) and $T_2$ (blue) on $b_x$. The temperature for this plot is $T=100\mbox{ mK}$, and the other parameters are listed in Sec.~\ref{subsubsec:Dep_T_Qubit11Tprime02Sprime}. The anticrossing is between $\ket{(1,1)T_-'}$ and $\ket{(0,2)S'}$.}
\label{fig:bx_dep}
\end{center}
\end{figure}

\begin{figure}[tb] 
\begin{center}
\includegraphics[width=\linewidth]{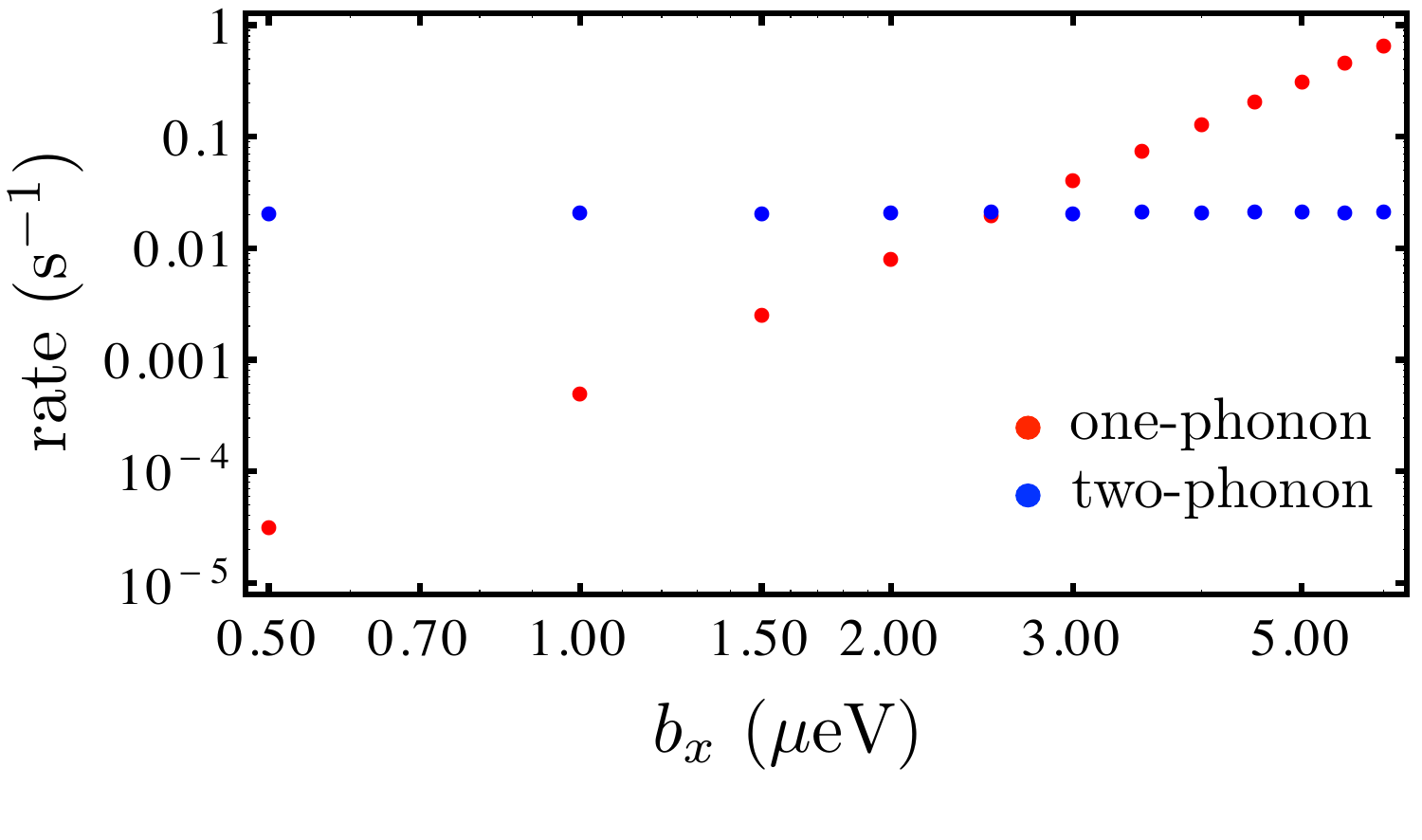}
\caption{The dependence of one-phonon ($\Gamma_2^{1p}$) and two-phonon ($\Gamma_2^{2p}$) components of the decoherence rate $1/T_2 = \Gamma^{1p}_2 + \Gamma^{2p}_2$ on $b_x$. The parameters are the same as in Fig.~\ref{fig:bx_dep}.}
\label{fig:bx_dep_rates}
\end{center}
\end{figure}

As the magnetic field gradient is determined by the design of the experimental setup, we also plot the dependence of $T_1$ and $T_2$ on $b_x$, see Fig.~\ref{fig:bx_dep}. The parameter values  we used are the same as for Fig.~\ref{fig:STminus_T} and $T=100\mbox{ mK}$. Here again $T_2\simeq 2T_1$. We can see a plateau up to $b_x\sim 2\mbox{ $\mu$eV}$ and then a decay for both $T_1$ and $T_2$. To explain this behavior we study the dependence of $\Gamma_2^{1p}$ and $\Gamma_2^{2p}$ on $b_x$, see Fig.~\ref{fig:bx_dep_rates}. Considering a fit function of the form $Y = C_i + C_{i+1} X^k$, as already used in Sec.~\ref{subsubsec:Dep_T_Qubit11Tprime02Sprime}, it turns out that $\Gamma_2^{1p} \simeq C_3+C_4 b_x^4$, where $C_3$ and $C_4$ are constants. In contrast to $\Gamma_2^{1p}$, the rate $\Gamma_2^{2p}$ does not change noticeably with $b_x$, and we will comment on this using a simple model in Sec.~\ref{subsubsec:SimpleModel}. Consequently, as for smaller $b_x$ the rate $\Gamma_2^{2p}$ dominates, there is a plateau in the regime of small $b_x$. For $b_x > 3\mbox{ $\mu$eV}$ the rate $\Gamma_2^{1p}$ dominates, therefore both $T_1$ and $T_2$ decay.

\subsubsection{Proposed experiments to confirm the theory}
\label{subsubsec:Experiments_Qubit11Tprime02Sprime}

In Fig.~\ref{fig:STminus_T} we see that phonon-assisted relaxation and decoherence are slow compared to the ones usually reported for GaAs. Nevertheless, we found regimes where $T_1$ and $T_2$ are in the millisecond range, so that phonon-assisted relaxation and dephasing may dominate over other sources of decoherence in the sample. This provides an option to test our theory experimentally. We suggest to consider two cases: when the one-phonon process dominates and when the two-phonon process dominates. 

\begin{figure}[tb] 
\begin{center}
\includegraphics[width=\linewidth]{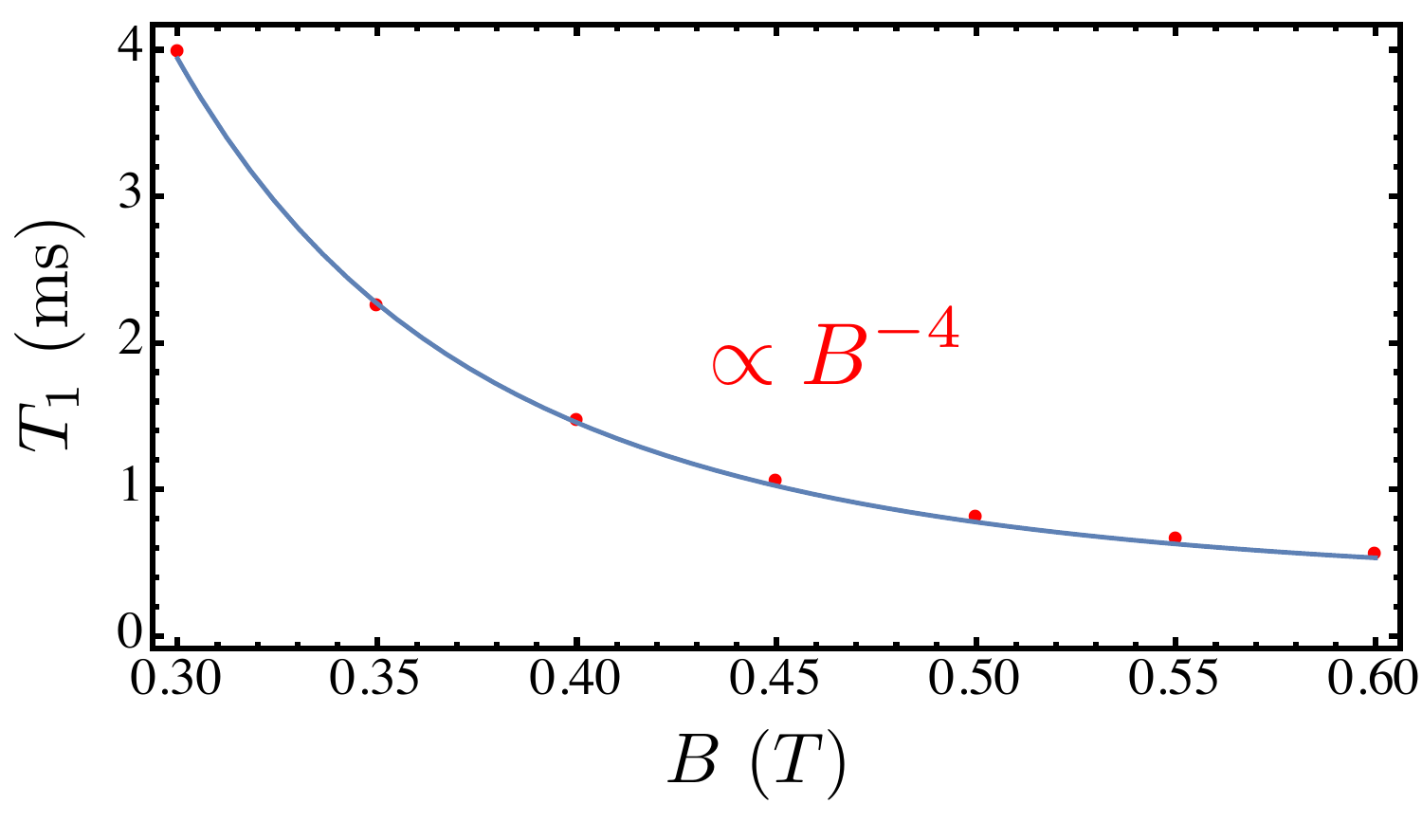}
\caption{The dependence of $T_1$ on the absolute value $B$ of the magnetic field for $T=100\mbox{ mK}$ and the parameters in the text. The blue line is a fit to the numerical results and shows a power-law decay~$\propto B^{-4}$. The detuning $\epsilon$ was chosen near (but not exactly at) the anticrossing  between $\ket{(1,1)T_-'}$ and $\ket{(0,2)S'}$. }
\label{fig:B_dep_oneph}
\end{center}
\end{figure}

To get the one-phonon process dominating, we use the following parameters: $b_x=10\mbox{ $\mu$eV}$, $\epsilon=1.173\mbox{ meV}$, $T=100\mbox{ mK}$, and the other parameters are the same as for Fig.~\ref{fig:STminus_T}. This means we have a similar spectrum as in Fig.~\ref{fig:S_qubit_Si} and stay in the region to the left from the marked anticrossing to have a large splitting between the qubit states, which increases the one-phonon relaxation rate. In Fig.~\ref{fig:B_dep_oneph} we plotted the dependence of $T_1$ on the applied magnetic field $B$. Here $T_1$ is mainly determined by the one-phonon process, therefore $T_1\simeq 1/\Gamma_1^{1p}$. The decay scales as $T_1\propto B^{-4}$. We note that we do not expect this power-law to be universal for all possible parameter values.

\begin{figure}[tb] 
\begin{center}
\includegraphics[width=\linewidth]{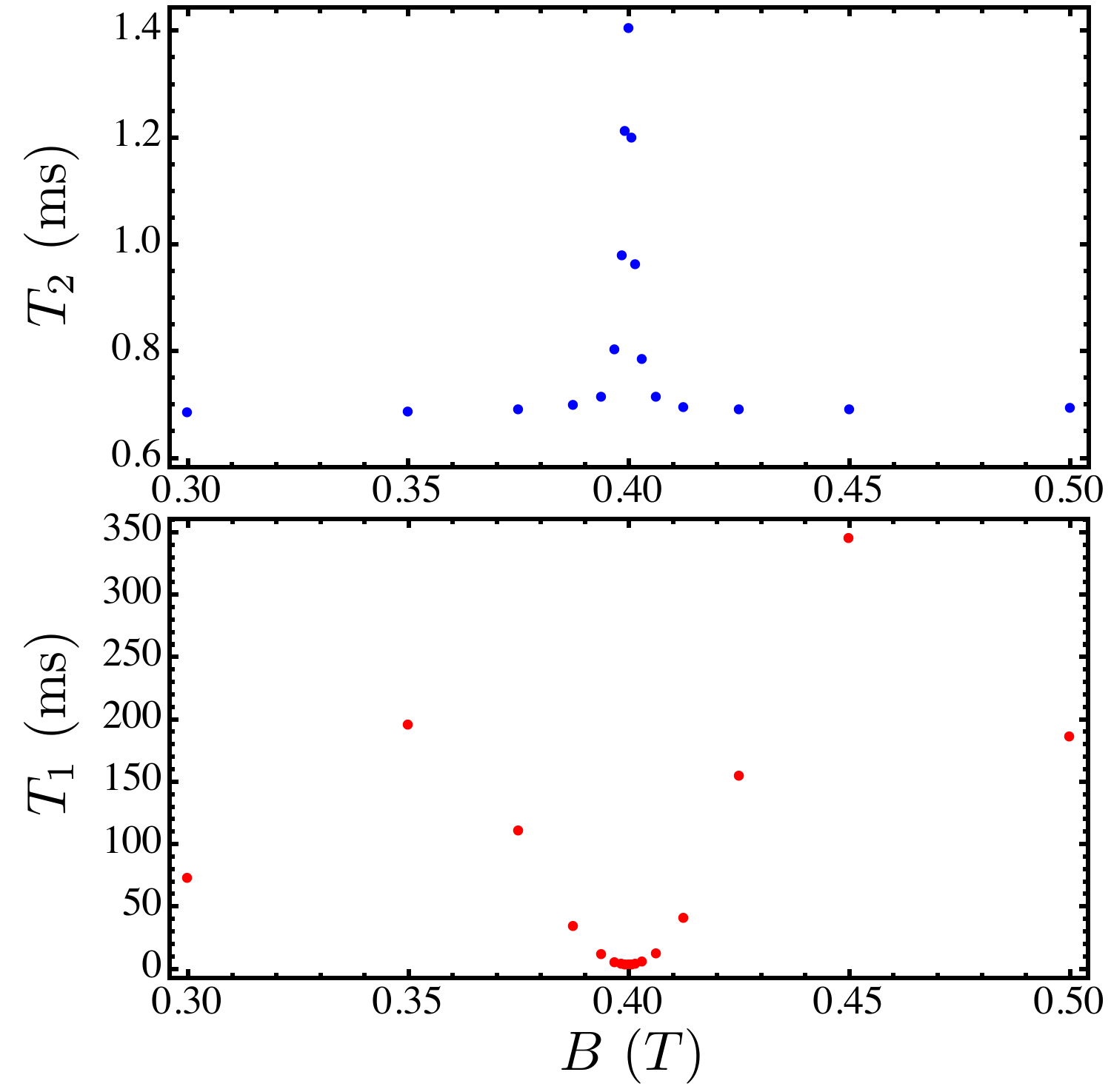}
\caption{The dependence of $T_2$ and $T_1$ on the absolute value $B$ of the magnetic field for $T=500\mbox{ mK}$ and the parameters in the text. The center of the anticrossing between $\ket{(1,1)T_-'}$ and $\ket{(0,2)S'}$ is at $B=0.4\mbox{ T}$. The peak of $T_2$ at $B=0.4\mbox{ T}$ is limited by $T_1$, i.e., $T_2\simeq 2T_1$, whereas the valleys at $B<0.395\mbox{ T}$, $B>0.405\mbox{ T}$ are predominantly determined by $T_\varphi$. In this figure, $T_2$ is due to two-phonon processes, one-phonon processes are negligible. For $T_1$, two-phonon processes dominate at $0.375\mbox{ T}<B<0.425\mbox{ T}$.}
\label{fig:B_dep_twoph}
\end{center}
\end{figure}

To test experimentally our theory of the two-phonon process, we suggest to change the magnetic field $B$ around the value where we are exactly in the center of the anticrossing, and to use rather small $b_x$. We need $b_x$ to be small enough, because at larger $b_x$ the one-phonon process starts to dominate as evident from Fig.~\ref{fig:bx_dep_rates}. Therefore, we use $b_x=1\mbox{ $\mu$eV}$, $T=500\mbox{ mK}$, and the other parameters as for Fig.~\ref{fig:STminus_T}. We plot the $B$ dependence of $T_2$ and $T_1$ in Fig.~\ref{fig:B_dep_twoph}. Here we see a sharp peak for $T_2$ at $B=0.4\mbox{ T}$, which is the center of the anticrossing between $\ket{(0,2)S'}$ and $\ket{(1,1)T'}$. Interestingly, dephasing is dominating for $B<0.395\mbox{ T}$, $B>0.405\mbox{ T}$, and the peak itself is limited by relaxation. The relaxation time $T_1$ is limited by two-phonon processes only at $0.375\mbox{ T}<B<0.425\mbox{ T}$. Apart from its usefulness for checking our theory of two-phonon processes the peak in $T_2$ (or dip in $T_1$) is a clear indication of the $\ket{(0,2)S'}$ and $\ket{(1,1)T_-'}$ anticrossing center, the point which is most interesting for spin qubit operation \cite{chesi:prb14, wong:prb15}.

Taking into account that the phonon-induced decoherence still allows for relatively long qubit lifetimes, we suppose that for the present-day samples the main source of decoherence in such a qubit will be charge noise, because the anticrossing region is quite narrow. However, as was discussed in Ref.~\onlinecite{chesi:prb14}, the charge noise can be substantially reduced using $t\gg E_Z$.

\subsubsection{Simple model for the qubit based on $\ket{(1,1)T_-'}$-$\ket{(0,2)S'}$} 
\label{subsubsec:SimpleModel}  

To analyze the results presented above, we propose to consider a simple model, which besides energy separation arguments discussed below is also justified by comparison with our numerical calculations. The states that are closest in energy to our $\ket{(1,1)T_-'}$-$\ket{(0,2)S'}$ qubit subspace are $\ket{(1,1)T_0}$ and $\ket{(1,1)S'}$. However, $\ket{(1,1)T_0}$ is decoupled from the qubit subspace when $b_B = 0$. Therefore we consider the Hamiltonian in the basis $\{\ket{(1,1)T_-}, \ket{(0,2)S}, \ket{(1,1)S}\}$:
\begin{eqnarray}
\tilde{H}&=&\begin{pmatrix} -E_Z& 0 & \frac{1}{2\sqrt{2}}b_x \\
0 & -\epsilon+U-V_-+\tilde{P} & -\sqrt{2} t +P_S  \\
\frac{1}{2\sqrt{2}}b_x & -\sqrt{2}t+P_S^\dagger  & V_+-V_- \\
\end{pmatrix} \nonumber \\
&&+H_{ph},
\label{eq:Hamiltonian_ST_Simple}
\end{eqnarray}
where $\tilde{P}=P_{SR}-P_T$. Our numerical calculation also showed that $P_S$ and $P_S^\dagger$ can be neglected, therefore  we will omit them in this subsection.

First of all we have to find the center of the $\ket{(1,1)T_-'}$-$\ket{(0,2)S'}$ anticrossing. For that we diagonalize the phonon-independent part of $\tilde{H}$ in the basis $\{\ket{(0,2)S},\ket{(1,1)S}\}$. This transformation is \cite{stepanenko:prb12}
\begin{equation}
U_1=\begin{pmatrix}1 & 0 & 0\\
0 &\cos{(\phi/2)} & -\sin{(\phi/2)}\\
0& \sin{(\phi/2)} & \cos{(\phi/2)}\\
\end{pmatrix},
\end{equation}
where
\begin{eqnarray}
\cos{\phi} &=& \frac{-U+V_++\epsilon}{\sqrt{8t^2+(-U+V_++\epsilon)^2}},\\
\sin{\phi} &=& \frac{2\sqrt{2}t}{\sqrt{8t^2+(-U+V_++\epsilon)^2}}.
\end{eqnarray}
Consequently, the matrix $U_1^\dagger \tilde{H} U_1$, with $U_1^\dagger$ as the conjugate transpose of $U_1$, corresponds to the Hamiltonian $\tilde{H}$ written in the basis $\{\ket{(1,1)T_-}$, $\ket{(0,2)S'}$, $\ket{(1,1)S'}\}$, if we set $b_x=0$. The anticrossing center is the point where the energy of $\ket{(1,1)T_-}$ is equal to the energy of $\ket{(0,2)S'}$ (with $b_x=0$ and $\tilde{P}=0$). From this condition we find the detuning $\epsilon$ at which the anticrossing occurs,
\begin{equation}
\epsilon=E_Z+U-V_--\frac{2t^2}{E_Z-V_-+V_+}.
\end{equation}
Assuming that $|b_x|$, $|t|$, and $\tilde{P}$ (may, e.g., be estimated via the expectation value of $\tilde{P}^2$) are much smaller than $\Delta=\sqrt{8t^2+(U-V_+-\epsilon)^2}$, we perform a Schrieffer-Wolff transformation up to the fourth order. The resulting Hamiltonian for the qubit and the phonons is then split into the part which does not contain phonons, the part with electron-phonon interaction, and $H_{ph}$. To simplify the analysis we apply to this Hamiltonian a unitary transformation $U_2$ which exactly diagonalizes the phonon-independent part,
\begin{equation}
U_2=\begin{pmatrix}\cos{(\Theta/2)} & -\sin{(\Theta/2)}\\
\sin{(\Theta/2)} & \cos{(\Theta/2)} \\ \end{pmatrix},
\end{equation}
where the angle $\Theta$ is defined as
\begin{eqnarray}
\cos{(\Theta/2)} &=& \frac{c}{\sqrt{c^2 + 1}},\\
\sin{(\Theta/2)} &=& \frac{1}{\sqrt{c^2 + 1}},
\end{eqnarray}
and
\begin{widetext}
\begin{equation}
c=\frac{4b_x\Delta^2(1+\cos{\phi})-b_x^3\cos^2({\frac{\phi}{2}})\cos{\phi}+\sqrt{\cos^4({\frac{\phi}{2}})b_x^2(b_x^2\cos{\phi}-8\Delta^2)^2+2\Delta^2(b_x^2-32\Delta^2+b_x^2\cos{\phi})^2\sin^2({\frac{\phi}{2}}})}{\sin({\frac{\phi}{2}})\sqrt{2}\Delta(b_x^2-32\Delta^2+b_x^2\cos{\phi})}.
\end{equation}

After all transformations our Hamiltonian is $H_q+H_{el-ph}(\tau)+H_{ph}$ as it was described in Sec.~\ref{subsec:BlochRedfieldTheory}. Therefore, in order to understand the results from Sec.~\ref{subsec:Tminus02S}, we present here the expression for $\delta B_{\tilde{x}}(\tau)$ which leads to relaxation and the one for $\delta B_{\tilde{z}}(\tau)$ which leads to dephasing,
\begin{eqnarray}
\label{eq:deltaBx}
\delta B_{\tilde{x}}(\tau) &=& \frac{1}{128\Delta^3} \biggl[\tilde{P}(\tau)\cos^2{\left[\frac{\phi}{2}\right]} \left(G_1\cos{\Theta}-G_2\sin{\Theta}\right)+\tilde{P}^2(\tau)\left(G_3\cos{\Theta}-G_4\sin{\Theta}\right)\biggr],\\
\label{eq:deltaBz}
\delta B_{\tilde{z}}(\tau) &=& \frac{1}{128\Delta^3} \biggl[\tilde{P}(\tau)\cos^2{\left[\frac{\phi}{2}\right]} \left(G_1\sin{\Theta}+G_2\cos{\Theta}\right)+\tilde{P}^2(\tau)\left(G_3\sin{\Theta}+G_4\cos{\Theta}\right)\biggr].
\end{eqnarray}
We note that $\delta B_{\tilde{y}}(\tau)=0$, and we introduced
\begin{eqnarray}
G_1 &=& \sqrt{2}b_x \sin{\left[\frac{\phi}{2}\right]} \left(b_x^2+32\Delta^2-7b_x^2\cos{\phi}\right) , \\
G_2 &=& 4\Delta \left(b_x^2-16\Delta^2-b_x^2\cos{\phi} \right) , \\
G_3 &=& 2\sqrt{2}b_x\sin{\phi}\cos{\left[\frac{\phi}{2}\right]} \Bigl( 5\tilde{P}(\tau) - 4\Delta+5\tilde{P}(\tau)\cos{[2\phi]} +[8\Delta-6\tilde{P}(\tau)]\cos{\phi} \Bigr) , \\
G_4 &=& 16\Delta\sin^2{\phi} \left(\Delta+\tilde{P}(\tau)\cos{\phi}\right)
\end{eqnarray}
for convenience. Using Eqs.~\eqref{eq:deltaBx} and \eqref{eq:deltaBz} we get the expressions for $\delta B_{\tilde{x}}(0)\delta B_{\tilde{x}}(\tau)$ and $\delta B_{\tilde{z}}(0)\delta B_{\tilde{z}}(\tau)$. To simplify them we use the fact that $\Theta\simeq\pi/2$ and $\Delta \gg |b_x|$, and get
\begin{eqnarray}
\label{eq:deltaBxdeltaBx}
&&\delta B_{\tilde{x}}(0)\delta B_{\tilde{x}}(\tau)\simeq\nonumber\\
&&\frac{1}{64\Delta^2}\left[16\cos^4{\left[\frac{\phi}{2}\right]}\Delta^2\tilde{P}(0)\tilde{P}(\tau)-4\cos^2{\left[\frac{\phi}{2}\right]}\cos{\phi}\sin^2{\phi}\big(\tilde{P}^3(0)\tilde{P}(\tau)+\tilde{P}(0)\tilde{P}^3(\tau)\big)+\sin^4{\phi}\tilde{P}^2(0)\tilde{P}^2(\tau)\right], \ \ \ \ \ \ \ \\
\label{eq:deltaBzdeltaBz}
&&\delta B_{\tilde{z}}(0)\delta B_{\tilde{z}}(\tau)\simeq\nonumber\\
&&\frac{1}{32\Delta^4}\cos^4{\left[\frac{\phi}{2}\right]}\sin^2{\left[\frac{\phi}{2}\right]}b_x^2[\cos{\phi}(5\cos{\phi}-3)(\tilde{P}^3(0)\tilde{P}(\tau)+\tilde{P}(0)\tilde{P}^3(\tau))+(2\cos{\phi}-1)^2\tilde{P}^2(0)\tilde{P}^2(\tau)].
\end{eqnarray}
\end{widetext}
The first term in the brackets in Eq.~\eqref{eq:deltaBxdeltaBx} is responsible for a one-phonon process, and the rest for two-phonon processes. We note that in the numerical calculations in this work we neglected terms of the type $\tilde{P}^3(0)\tilde{P}(\tau)$ and $\tilde{P}(0)\tilde{P}^3(\tau)$. The relaxation mechanism that results from these terms can be interpreted as a higher-order correction to the standard one-phonon process. In the presence of phonons which are neither absorbed nor emitted, one phonon matches the Zeeman energy and ensures energy conservation. Furthermore, analogous to the standard terms of a one-phonon process, such terms do not contribute to dephasing at all \cite{kornich:prb14}. 

The coefficient of $\tilde{P}^2(0)\tilde{P}^2(\tau)$ in Eq.~\eqref{eq:deltaBxdeltaBx} is more than $1000$ times larger than the coefficient  of the same term in Eq.~\eqref{eq:deltaBzdeltaBz} for the parameter values we used for Fig.~\ref{fig:STminus_T}. This suggests that two-phonon-based dephasing is negligibly small compared to two-phonon-based relaxation, and explains why in Fig.~\ref{fig:STminus_T} we have $T_2\simeq 2T_1$. 
Qualitatively, the presented relaxation via two-phonon processes can be understood as follows. At the anticrossing, the eigenstates of the qubit Hamiltonian are approximately $\left[\ket{(0,2)S'} \pm \ket{(1,1)T_-}\right] /\sqrt{2}$. Two-phonon Raman processes \cite{mccumber:jap63, yen:pr64, altner:prb96, meltzer:book05, roszak:prb09} based on the singlet states of the biased DQD \cite{kornich:prb14, kornich:prb14er} efficiently shift the energy of $\ket{(0,2)S'}$, which corresponds to a transverse coupling in the qubit subspace and therefore leads to relaxation.

From Eqs.~\eqref{eq:deltaBxdeltaBx} and \eqref{eq:deltaBzdeltaBz} it is evident that the dephasing part depends on $b_x$ strongly, whereas for relaxation $b_x$ enters only with $\Delta_{ST}$ in $J_{\tilde{x}\tilde{x}}^+(\Delta_{ST})$. The explicit expressions for $\int_{-\infty}^\infty\cos{(\Delta_{ST} \tau/\hbar)}\langle\tilde{P}^2(0)\tilde{P}^2(\tau)\rangle d\tau$ show that $\Delta_{ST}$ enters in the dominating terms as $q+\Delta_{ST}/(\hbar v_l)$, where $\hbar q$ is the momentum of a phonon.  The integrals over $q$ (continuum limit) can simply be performed from 0 to $\infty$  and converge because of Bose-Einstein terms or because of the Gaussian terms that result when integrating out the spatial dependence of the electron wave functions combined with oscillations of type $e^{i \bm{q}\cdot\bm{r}}$. We note that these Gaussian terms have decayed when the phonon wavelength is (much) smaller than the size of a QD \cite{golovach:prl04, meunier:prl07, golovach:prb08}.  The main contribution to the rates is provided by the part of the integrals with $q\gg \Delta_{ST}/(\hbar v_l)$ within the range of parameters used for Figs.~\ref{fig:STminus_T} and \ref{fig:bx_dep}. Therefore, for the two-phonon relaxation process, the effect of $b_x$ is negligible, which is seen in Fig.~\ref{fig:bx_dep_rates}.

\subsection{The qubit based on $\ket{(1,1)T_-'}$-$\ket{(1,1)S'}$}
\label{subsec:Tminus11S}

\begin{figure}[tb] 
\begin{center}
\includegraphics[width=\linewidth]{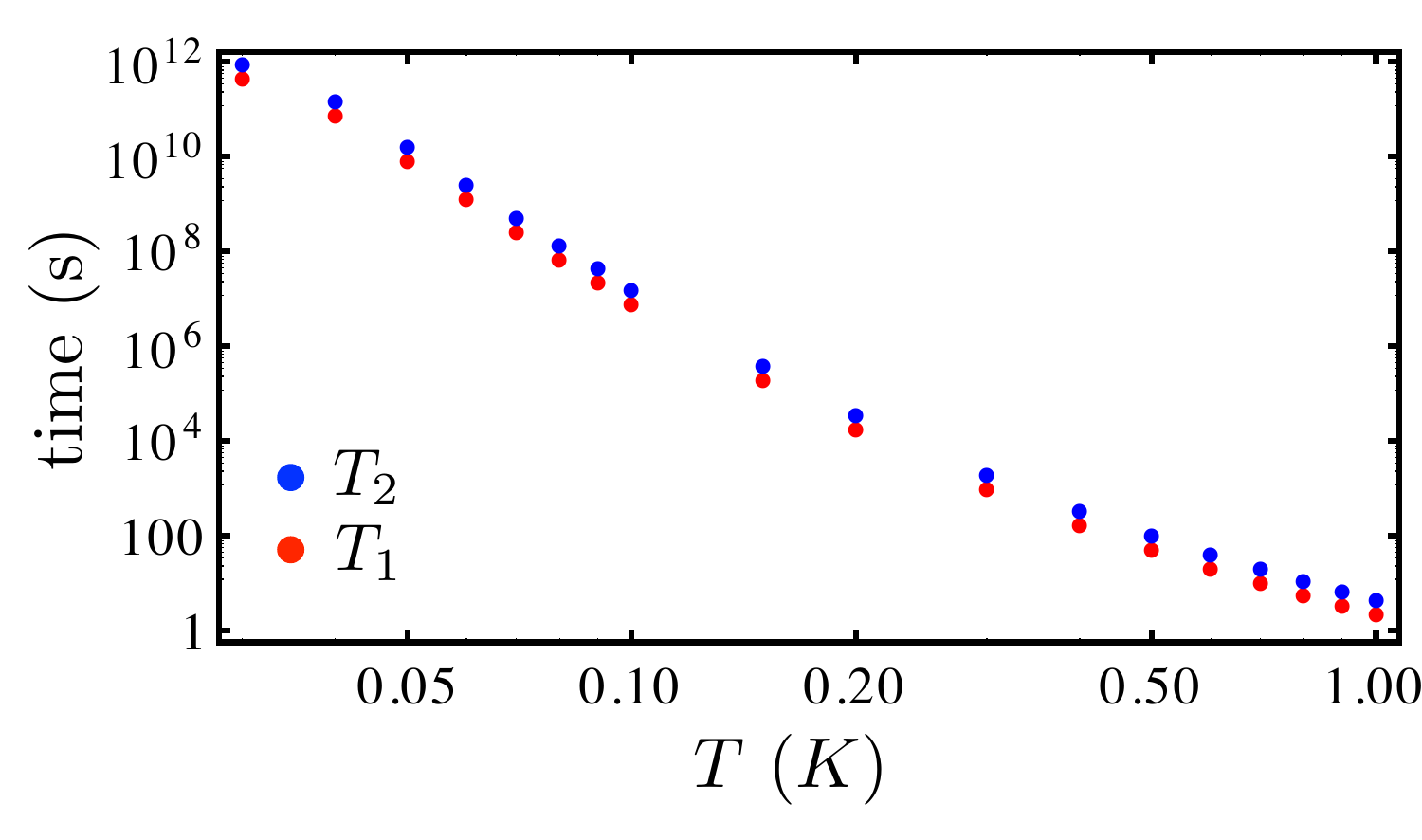}
\caption{The dependence of relaxation time $T_1$ (red) and $T_2$ (blue) on temperature. The anticrossing is between $\ket{(1,1)T_-'}$ and $\ket{(1,1)S'}$. For the parameters, see Sec.~\ref{subsec:Tminus11S}. }
\label{fig:S11Tminus_T}
\end{center}
\end{figure}

\begin{figure}[tb] 
\begin{center}
\includegraphics[width=\linewidth]{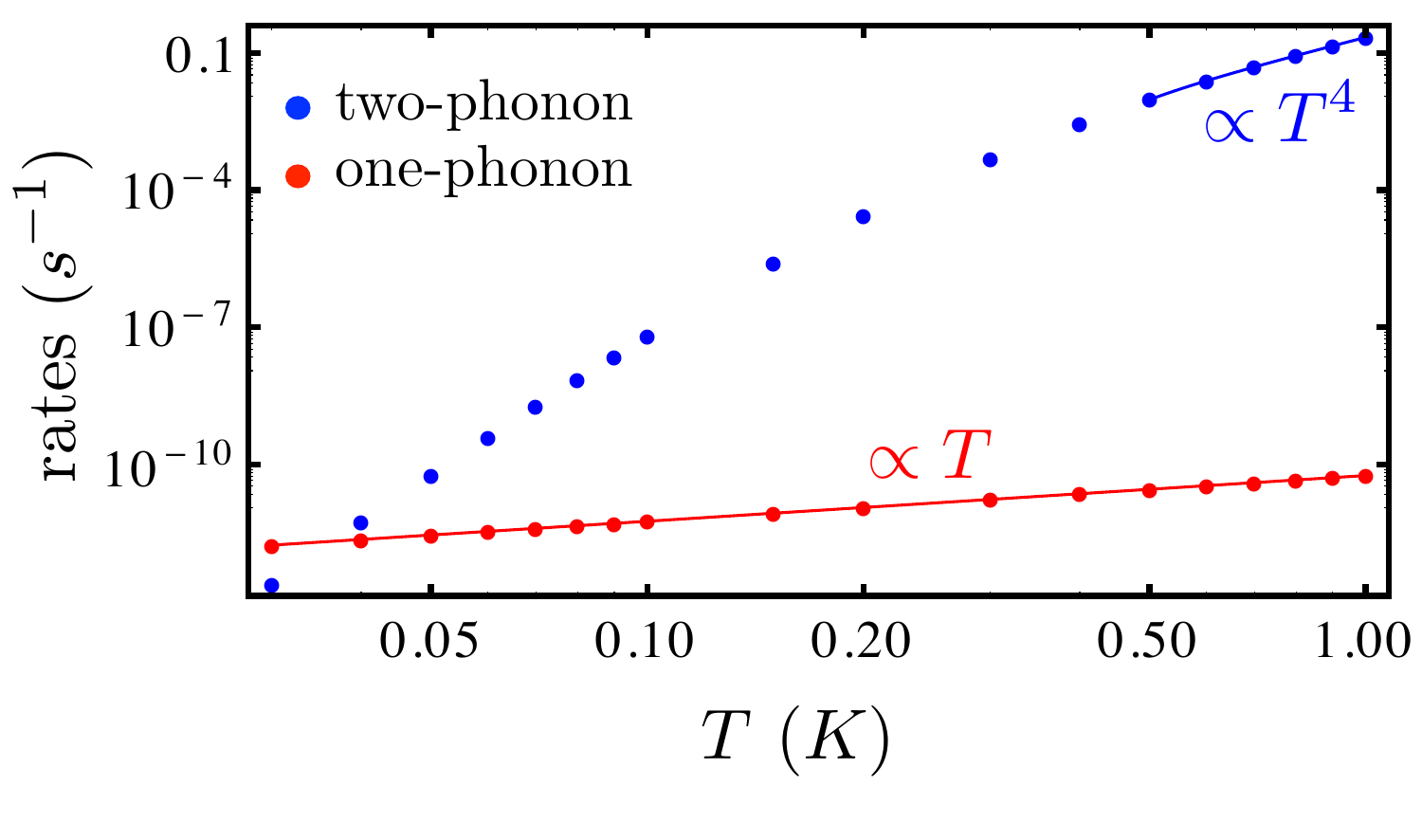}
\caption{The dependence of one-phonon ($\Gamma_2^{1p}$) and two-phonon ($\Gamma_2^{2p}$) components of the decoherence rate on temperature. The anticrossing is between $\ket{(1,1)T_-'}$ and $\ket{(1,1)S'}$. The straight lines obey the shown power-laws and are fits to our numerical results. The parameters are the same as in Fig.~\ref{fig:S11Tminus_T}.}
\label{fig:S11Tminus_rates_T}
\end{center}
\end{figure}

Now let us consider the case where the qubit is based on the anticrossing $\ket{(1,1)T_-'}$-$\ket{(1,1)S'}$, as shown in Fig.~\ref{fig:F_qubit_Si}. For this, we plot the temperature dependence of $T_1$ and $T_2$ (see Fig.~\ref{fig:S11Tminus_T}) using the following parameters: $B=4.5\mbox{ mT}$, $t=10\mbox{ $\mu$eV}$, $V_+=40\mbox{ $\mu$eV}$, $V_-=39.99\mbox{ $\mu$eV}$, $U=1.2\mbox{ meV}$, $b_x=0.1\mbox{ $\mu$eV}$, $L=150\mbox{ nm}$, $l_c=42.7\mbox{ nm}$, $\epsilon=0.68737\mbox{ meV}$. We see that again $T_2\simeq 2T_1$. In Fig.~\ref{fig:S11Tminus_rates_T} we plotted the dependence of $\Gamma_2^{1p}$ and $\Gamma_2^{2p}$ on temperature. The transition where the two-phonon process starts to dominate over the one-phonon process is now at lower temperature than for the case plotted in Fig.~\ref{fig:STminus_rates_T}. We note that at $0.03\mbox{ K}<T<0.08\mbox{ K}$, a fit yields $\Gamma_2^{2p} \simeq C_5+C_6T^{10}$ for the two-phonon process rate, where $C_5$, $C_6$ are constants. 

Remarkably, from Fig.~\ref{fig:S11Tminus_T} it follows that phonon-induced relaxation and decoherence are extremely slow. However, as we noted before, we neglected the effect of SOI in this calculation. When $b_x$ is very small, it can be that SOI effects are noticeable. Let us assume there is Rashba SOI in our sample. Then, for the values of $b_x$, $E_Z$, $L$, and $l_c$ we use in this subsection, the Rashba SOI length must be $l_R\simeq 1.6\mbox{ $\mu$m}$ for $2 \Omega$ to be of the same absolute value as $b_x$. We note that in GaAs/AlGaAs heterostructures $l_R$ of the order of $1\mbox{ $\mu$m}$ has been reported \cite{hanson:rmp07}. Although we are not aware of precise values for the SOI of electrons in lateral Si/SiGe-based QDs, we expect it to be weaker ($l_R$ longer) than in GaAs/AlGaAs.

We note that the $\ket{(1,1)T_-'}$-$\ket{(1,1)S'}$-type qubit is also robust against charge noise, because the qubit is operated at the ``sweet spot'', where $\partial \Delta_{ST}/\partial \epsilon\simeq 0$, and the anticrossing region is wide \cite{wong:prb15}.

\section{$S$-$T_0$ qubit}
\label{sec:ST0Qubit}

In this section we consider the qubit based on $\ket{(1,1)S'}$-$\ket{(1,1)T_0'}$. There are two cases which we are interested in. The first one is the region of large detuning, where $\ket{(0,2)S'}$ is close to the qubit subspace. The second one is the zero-detuning case, where we have to take excited orbital states into account. We already considered these cases in our previous work on DQDs in GaAs/AlGaAs heterostructures, Ref.~\onlinecite{kornich:prb14}. Here we consider a DQD in a SiGe/Si/SiGe quantum well and present the dependence of $T_1$ and $T_2$ on different quantities which were not studied in our previous work.

\subsection{Large detuning}
\label{subsec:LargeDetuning}

In this subsection we consider the region near the anticrossing of $\ket{(1,1)S'}$ and $\ket{(0,2)S'}$, where the state $\ket{(0,2)S'}$ is sufficiently closer to the qubit subspace than states with excited orbital parts, so that the latter can be omitted. We use the Hamiltonian from Eq.~\eqref{eq:HamiltonianSmall} and calculate $T_1$ and $T_2$ using the theory described in Sec.~\ref{subsec:BlochRedfieldTheory}. At the end of this subsection we present a simple analytic model and discuss our numerical results.

\subsubsection{Dependence on the magnetic field gradient}
\label{subsubsec:Dep_bB}

\begin{figure}[tb] 
\begin{center}
\includegraphics[width=\linewidth]{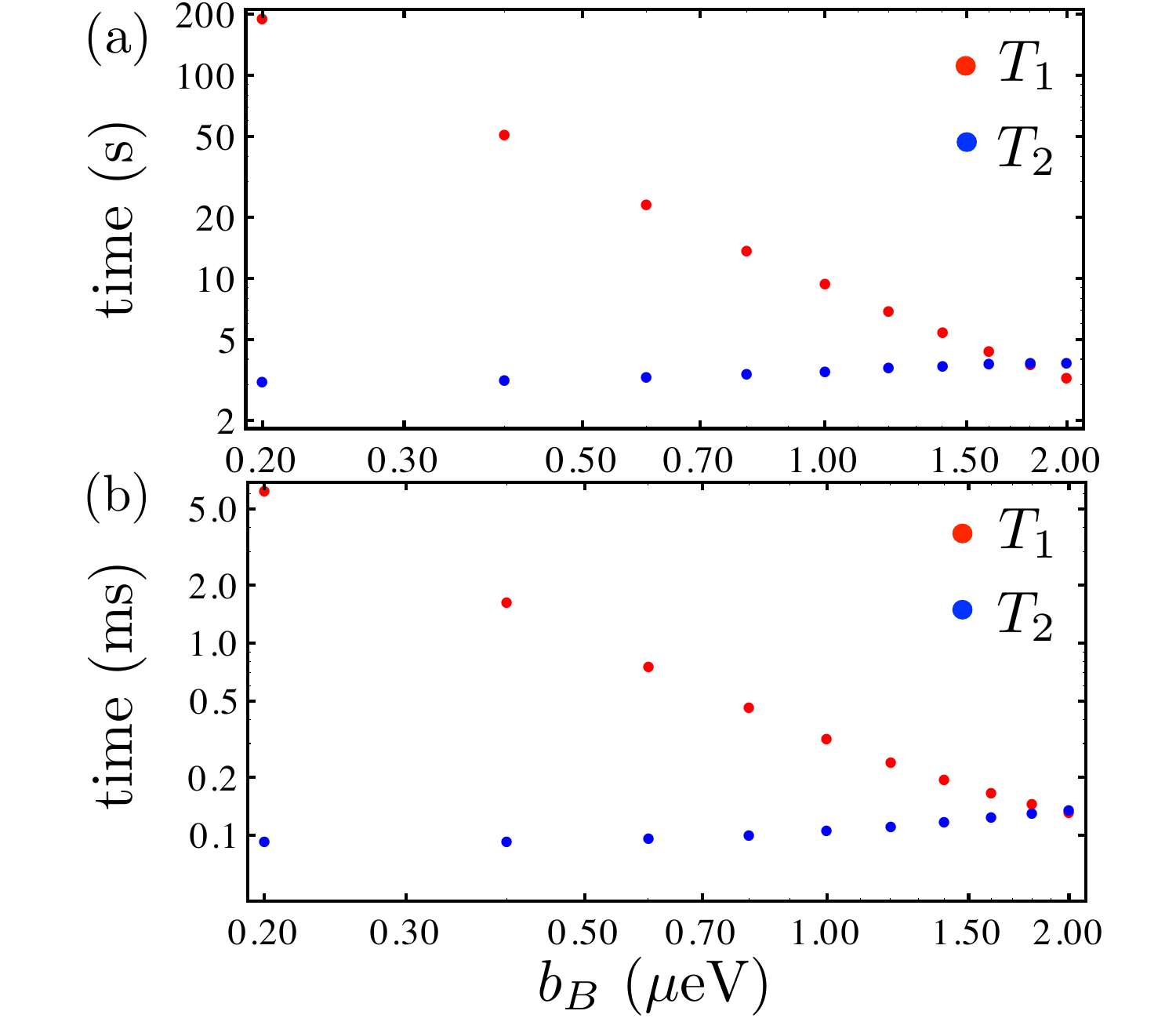}
\caption{(a) The dependence of $T_1$ (red) and $T_2$ (blue) on $b_B$ for a $S$-$T_0$ qubit at large detuning. The temperature is $T=100\mbox{ mK}$. The decoherence time $T_2$ is slightly increasing (except for the last point) with increasing $b_B$, whereas the relaxation time decreases drastically. (b) The same dependence as in panel a, but at $T=500\mbox{ mK}$. Other parameters are given in Sec.~\ref{subsubsec:Dep_bB}.}
\label{fig:dbb_dep}
\end{center}
\end{figure}

We study the dependence of $T_1$ and $T_2$ on the energy $b_B$ associated with the magnetic field gradient. For Fig.~\ref{fig:dbb_dep} we used $B=0.4\mbox{ T}$, $t=4\mbox{ $\mu$eV}$, $V_+=40\mbox{ $\mu$eV}$, $V_-=39.99\mbox{ $\mu$eV}$, $U=1.2\mbox{ meV}$, $b_x=0$, $L=150\mbox{ nm}$, $l_c=42.7\mbox{ nm}$, and $\epsilon=1.144\mbox{ meV}$. The chosen confinement length corresponds to the level splitting $\Delta E=200\mbox{ $\mu$eV}$, which allows us to neglect the effect of the excited states compared to $\ket{(0,2)S'}$ due to the large energy gap. As we took $b_x=0$, we consider non-zero Rashba SOI. The Rashba SOI length we use is quite short, $l_R=2\mbox{ $\mu$m}$, and we take $\eta = 0$ to make the effect of SOI maximal [see Eq.~\eqref{eq:SOI}], resulting in $\Omega = -3.48\mbox{ $\mu$eV}$. However, our numerical calculation of the qubit lifetimes showed that the effect of SOI in this regime of large detuning, even with a rather small $l_R$ and $\eta=0$, is negligible. For the parameters described above and the range of $b_B$ in Fig.~\ref{fig:dbb_dep}, the resulting $\Delta_{ST}$ is in the range $1.8\mbox{ $\mu$eV}<\Delta_{ST}<2.6\mbox{ $\mu$eV}$.

From Fig.~\ref{fig:dbb_dep} we see that the behavior of $T_1$ and $T_2$ is similar at $T=100\mbox{ mK}$ and $T=500\mbox{ mK}$. We note that in contrast to our previous work for GaAs QDs \cite{kornich:prb14}, the relation $T_\varphi \ll T_1$ does not hold for the whole parameter range. For $b_B<0.8\mbox{ $\mu$eV}$ the pure dephasing part $T_{\varphi}$ dominates over $T_1$, bringing $T_2$ to much lower values than $T_1$. However, as the magnetic field gradient enhances relaxation processes strongly, $T_1$ decays rapidly with $b_B$, becoming of the order of $T_2$ and even $T_1 < T_2$ for $b_B > 1.6\mbox{ $\mu$eV}$. The strong dependence of relaxation on $b_B$ is easy to understand from the Hamiltonian [Eq.~\eqref{eq:HamiltonianSmall}], because $b_B/2$ is the off-diagonal term between $\ket{(1,1)S}$ and $\ket{(1,1)T_0}$ and the only term that couples $\ket{(1,1)T_0}$ to other states. This means that relaxation occurs only in case $b_B\neq 0$ and strongly depends on the value of $b_B$. We obtain $T_1\propto b_B^{-2}$ for $b_B<1\mbox{ $\mu$eV}$ in Fig.~\ref{fig:dbb_dep} (both a and b).

\subsubsection{Dependence on temperature}
\label{subsubsec:TemperatureDependence}

\begin{figure}[tb] 
\begin{center}
\includegraphics[width=\linewidth]{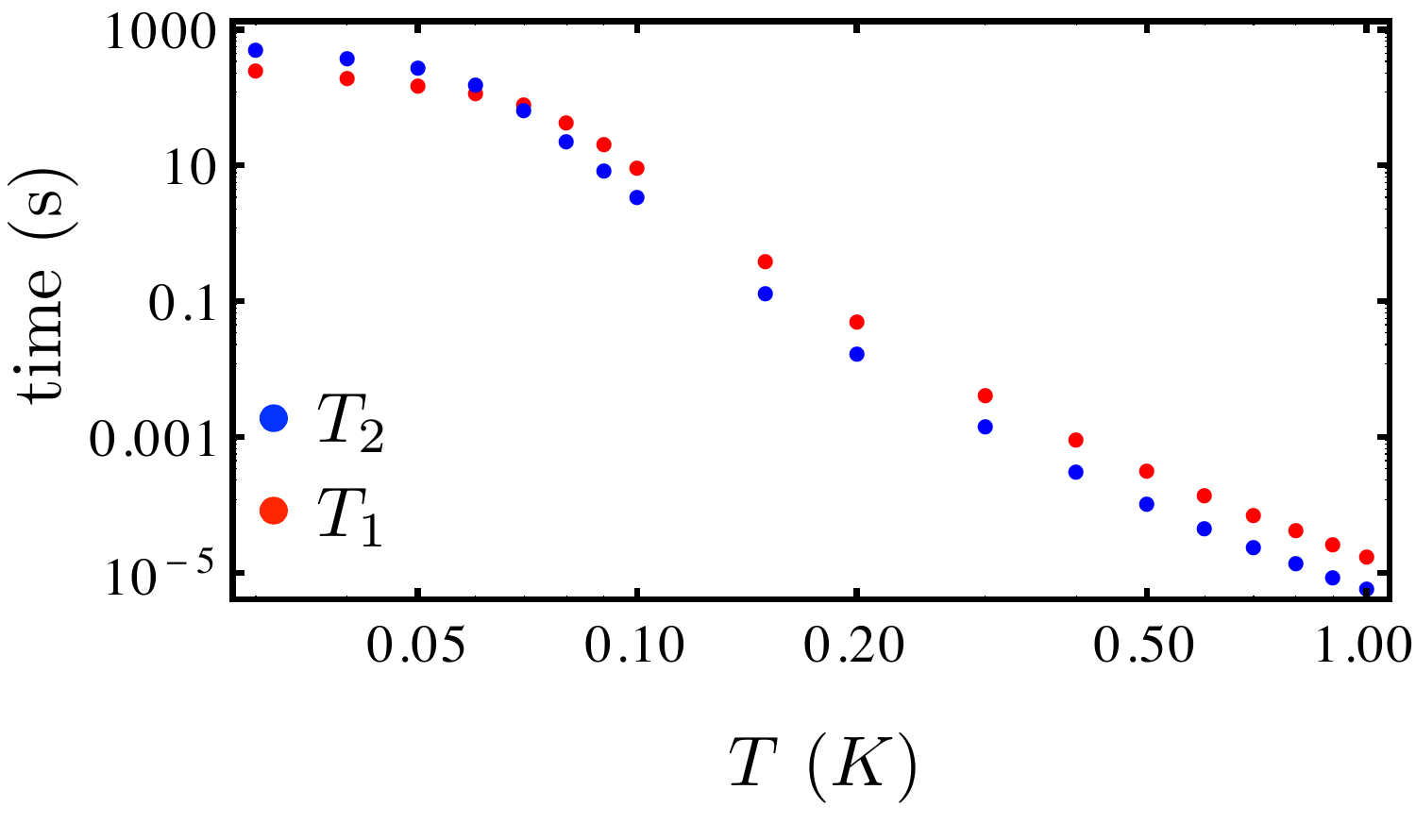}
\caption{The dependence of $T_2$ (blue) and $T_1$ (red) on temperature~$T$ for a $S$-$T_0$ qubit at large detuning $\epsilon$. }
\label{fig:temp_dep_large_det}
\end{center}
\end{figure}

We also show the temperature dependence of $T_1$ and $T_2$, see Fig.~\ref{fig:temp_dep_large_det}. For this plot we used $b_B=-1\mbox{ $\mu$eV}$ and otherwise the same parameters as for Fig.~\ref{fig:dbb_dep}. The splitting between the qubit states $\ket{(1,1)S'}$ and $\ket{(1,1)T_0'}$ is $\Delta_{ST}\simeq 2\mbox{ $\mu$eV}$. From Fig.~\ref{fig:temp_dep_large_det} we see that both $T_2$ and $T_1$, as expected, decrease with temperature. At very low temperatures, i.e., $T<0.06\mbox{ K}$, $T_2>T_1$. Then, however, $T_2$ decays faster than $T_1$. For $0.5\mbox{ K}\leq T\leq 1\mbox{ K}$, their power-laws are the same, $\propto T^{-4}$.  

\begin{figure}[tb] 
\begin{center}
\includegraphics[width=\linewidth]{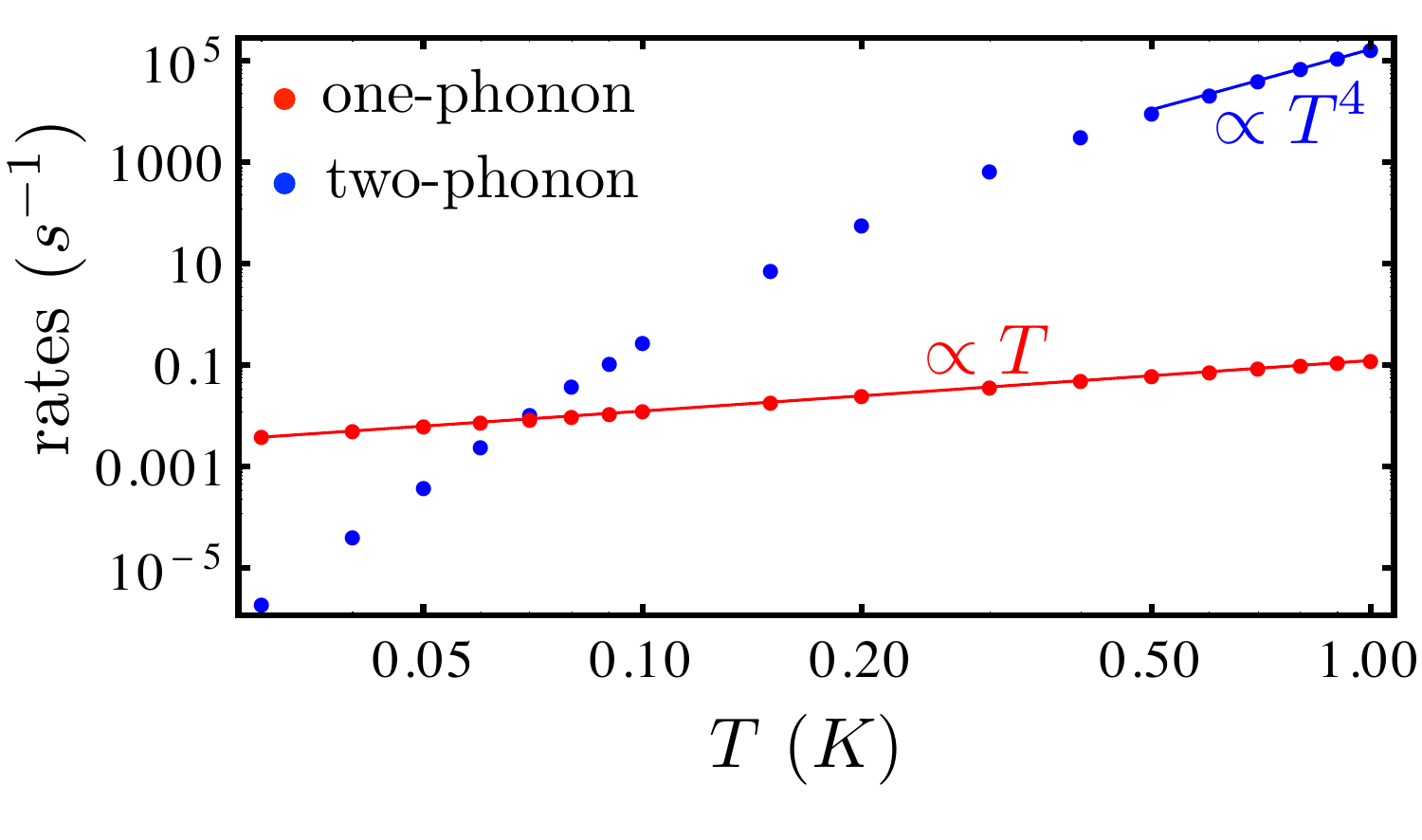}
\caption{The dependence of one-phonon ($\Gamma_2^{1p}$) and two-phonon ($\Gamma_2^{2p}$) components of the decoherence rate on temperature $T$. The straight lines are fits to our numerical results and obey the indicated power-laws. The parameters are the same as for Fig.~\ref{fig:temp_dep_large_det} and are provided in Sec.~\ref{subsec:LargeDetuning}. }
\label{fig:temp_dep_large_ep}
\end{center}
\end{figure}

To understand why $T_2$ (similarly for $T_1$) decays so slowly for $T<0.06\mbox{ K}$ and then faster, we plot the temperature dependence of $\Gamma_2^{1p}$ and $\Gamma^{2p}_2$ (see Fig.~\ref{fig:temp_dep_large_ep}). Here we see that for $T\leq 0.05\mbox{ K}$ the one-phonon process dominates and $\Gamma_2^{1p}\propto T$, which gives a slow decay of $T_1$ and $T_2$ with temperature. The origin of this dependence of $\Gamma_2^{1p}$ is the same as the one explained in Sec.~\ref{subsubsec:Dep_T_Qubit11Tprime02Sprime}. For temperatures $T\geq 0.1\mbox{ K}$ the two-phonon process dominates. Therefore, as $\Gamma^{2p}_2\propto T^4$ for $0.5\mbox{ K}\leq T\leq 1\mbox{ K}$, we see the same power-law for $1/T_2$. With a similar analysis for $T_1$, we find that for $0.5\mbox{ K}\leq T\leq 1\mbox{ K}$ also $1/T_1 \propto T^4$ due to two-phonon processes. Since we have a rather large $b_B$, the dephasing part $T_\varphi$ is of the same order as $T_1$, as was shown in Sec.~\ref{subsubsec:Dep_bB}. The reason for choosing here a large $b_B$ is the reported values for applied magnetic field gradients in experiments with micromagnets \cite{chesi:prb14,wu:pnas14, kawakami:nnano14, shin:prl10}.

\subsubsection{Dependence on detuning}
\label{subsubsec:ST0LargeDetDepOnDet}

\begin{figure}[tb] 
\begin{center}
\includegraphics[width=\linewidth]{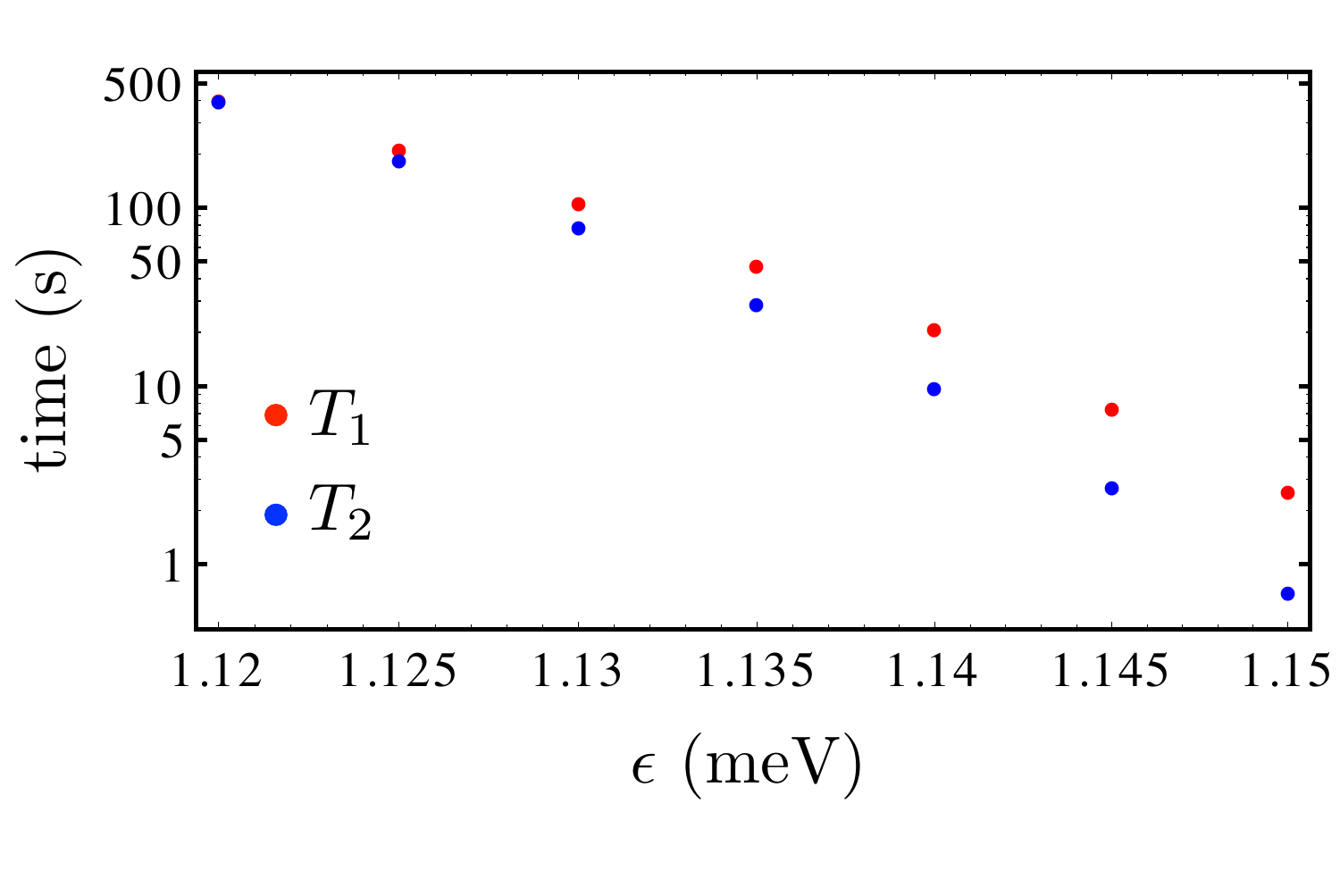}
\caption{The dependence of $T_1$ (red) and $T_2$ (blue) on detuning $\epsilon$ for a $S$-$T_0$ qubit at $T = 100\mbox{ mK}$. Details are described in the text.}
\label{fig:detuning_dep}
\end{center}
\end{figure}

Here we show that in the anticrossing region even small changes of $\epsilon$ affect both $T_1$ and $T_2$ strongly. For that we present the dependence of $T_1$ and $T_2$ on $\epsilon$, see Fig.~\ref{fig:detuning_dep}, where we used the same parameters as for Fig.~\ref{fig:dbb_dep} and took $b_B=-1\mbox{ $\mu$eV}$ and $T=100\mbox{ mK}$. For the range of $\epsilon$ shown in Fig.~\ref{fig:detuning_dep} the splitting $\Delta_{ST}$ takes values in the range $1.3\mbox{ $\mu$eV}<\Delta_{ST}<2.7\mbox{ $\mu$eV}$. We see that even though the change of $\epsilon$ is only  $30\mbox{ $\mu$eV}$, the relaxation time and decoherence time both change drastically. The main reason for this behavior is that in this region $\ket{(0,2)S'}$ very quickly drops in energy with $\epsilon$ and hence comes closer to the qubit subspace. 

\subsubsection{Dependence on tunnel coupling}

\begin{figure}[tb]
\begin{center}
\includegraphics[width=\linewidth]{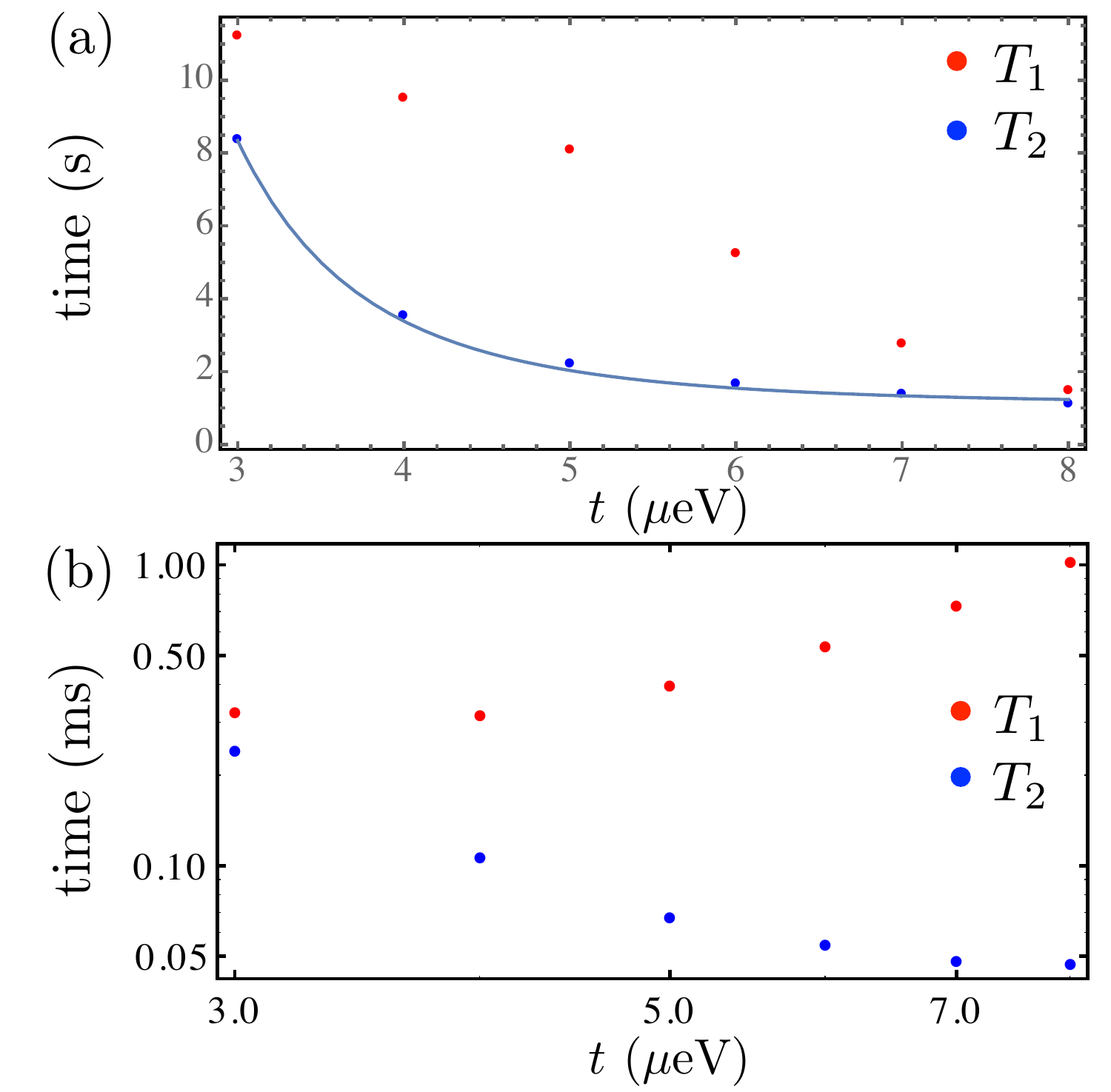}
\caption{(a) The dependence of $T_2$ (blue) and $T_1$ (red) on the tunnel coupling $t$ for a $S$-$T_0$ qubit in a biased DQD. Here the temperature is $100\mbox{ mK}$, for the other parameters see Sec.~\ref{subsec:LargeDetuning}. The blue line shows the fit with the function $T_2 = C_7 + C_8 t^{-4}$. Both $T_1$ and $T_2$ decay with~$t$. (b) The same dependence as in panel a, but at a higher temperature $T=500\mbox{ mK}$. Here we see that $T_1$ grows with $t$, in contrast to the case with $T=100\mbox{ mK}$ shown in panel~a.}
\label{fig:t_dep}
\end{center}
\end{figure} 

\begin{figure}[tb]
\begin{center}
\includegraphics[width=\linewidth]{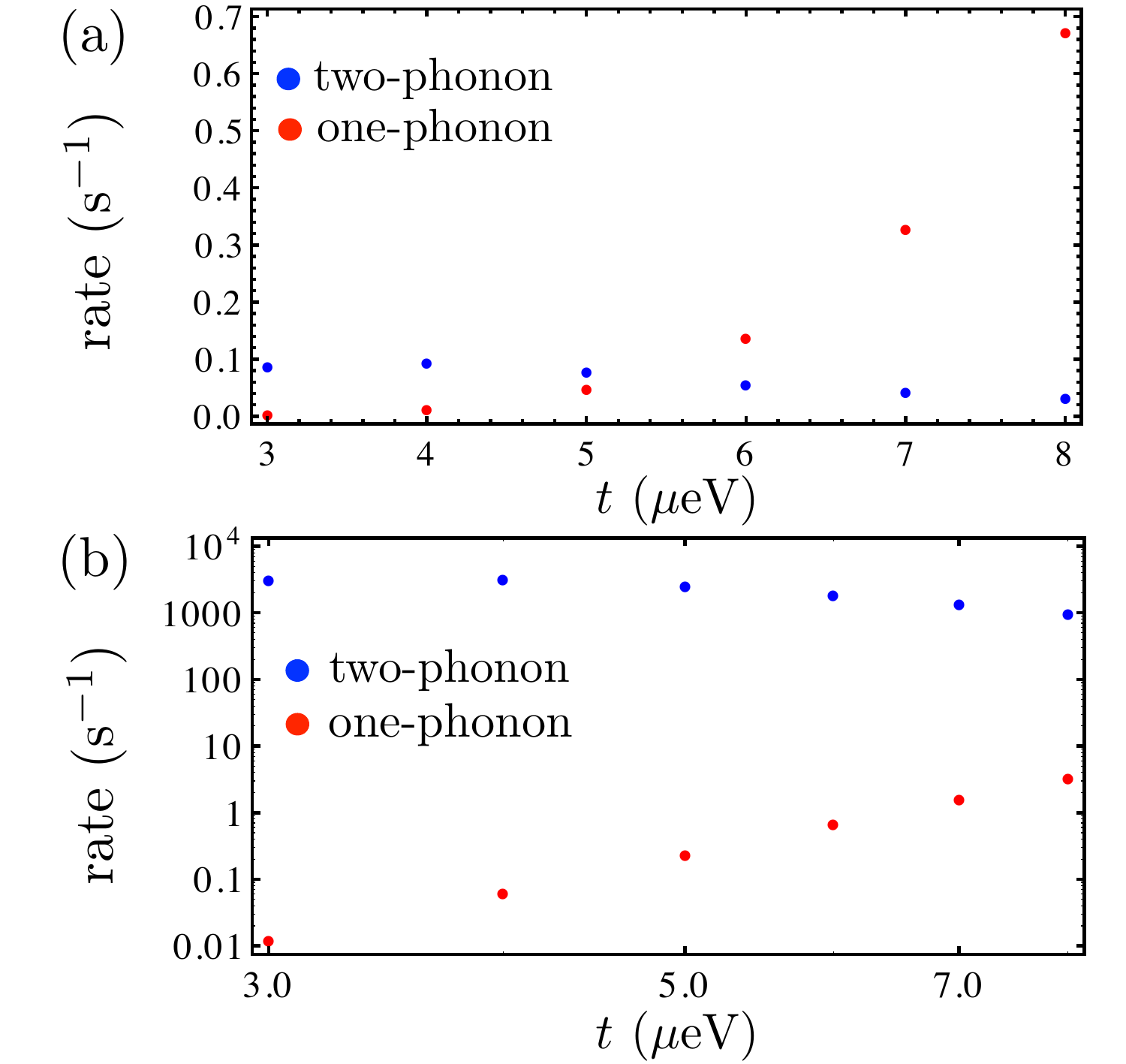}
\caption{The dependence of relaxation rates due to one-phonon processes ($\Gamma_1^{1p}$) and two-phonon processes ($\Gamma_1^{2p}$) on the tunnel coupling~$t$. The temperature is $100\mbox{ mK}$ in panel~a and $500\mbox{ mK}$ in panel~b. The parameters are the same as for Fig.~\ref{fig:t_dep}. }
\label{fig:rates_t_dep}
\end{center}
\end{figure} 

To find the optimal regime for qubit operation we present the dependence of $T_1$ and $T_2$ on the tunnel coupling $t$ between the dots, see Fig.~\ref{fig:t_dep}. For this calculation we used $b_B=-1\mbox{ $\mu$eV}$ and the other parameters as for Fig.~\ref{fig:dbb_dep}. The $S$-$T_0$ splitting changed with $3\mbox{ $\mu$eV} < t < 8\mbox{ $\mu$eV}$ in the interval $1.4\mbox{ $\mu$eV}<\Delta_{ST}<5.9\mbox{ $\mu$eV}$. For Fig.~\ref{fig:t_dep}a we used $T=100\mbox{ mK}$ and for Fig.~\ref{fig:t_dep}b $T=500\mbox{ mK}$. From Fig.~\ref{fig:t_dep}a we see that both $T_1$ and $T_2$ decay with $t$. However, the forms of their decays are different. The decay of $T_2$ reveals an approximate dependence $T_2 \simeq C_7 + C_8 t^{-4}$, where $C_7$ and $C_8$ are constants (the blue line in Fig.~\ref{fig:t_dep}a). In Fig.~\ref{fig:t_dep}b we see that $T_2$ decays with $t$. However, $T_1$ grows with $t$ for $t>4\mbox{ $\mu$eV}$. 

To understand this behavior of $T_1$ we plot the dependence of relaxation rates due to one-phonon ($\Gamma_1^{1p}$) and two-phonon processes ($\Gamma_1^{2p}$) on $t$ again for $100\mbox{ mK}$ and $500\mbox{ mK}$ (see Fig.~\ref{fig:rates_t_dep}). The rates satisfy
\begin{equation}
\frac{1}{T_1} = \Gamma_1=\Gamma_1^{1p}+\Gamma_1^{2p} .
\end{equation}
In both Figs.~\ref{fig:rates_t_dep}a ($100\mbox{ mK}$) and \ref{fig:rates_t_dep}b ($500\mbox{ mK}$) the one-phonon  rate grows with $t$, whereas the two-phonon  rate slowly decays at $t>4\mbox{ $\mu$eV}$. The difference in behavior of $T_1$ in Figs.~\ref{fig:t_dep}a and \ref{fig:t_dep}b arises from the fact that for the lower temperature, i.e., $100\mbox{ mK}$, for $t\geq 6\mbox{ $\mu$eV}$ the one-phonon relaxation rate dominates, which makes $T_1$ decrease with $t$. However, at larger temperature, $T=500\mbox{ mK}$, the two-phonon process starts to dominate (see Fig.~\ref{fig:rates_t_dep}b), which makes $T_1$ grow with~$t$.

\subsubsection{Simple model for the $S$-$T_0$ qubit at large detuning}
\label{subsec:SimpleModel2}

To understand the dependences on different parameters presented above, we consider a simple model. Similarly to the simple model of Ref. \onlinecite{kornich:prb14}, we consider the Hamiltonian
\begin{equation}
\tilde{H}=\begin{pmatrix}
0 && \frac{b_B}{2} && 0\\
\frac{b_B}{2} && V_+-V_- && -\sqrt{2}t \\
0 && -\sqrt{2}t && -\epsilon+U-V_-+\tilde{P}
\end{pmatrix}+H_{ph} 
\end{equation}
in the basis $\{\ket{(1,1)T_0}$, $\ket{(1,1)S}$, $\ket{(0,2)S}\}$, because the effect of $\ket{(0,2)S}$ on the qubit lifetimes is dominating. Here $\tilde{P}=P_{SR}-P_T$, and we note that the electron-phonon interaction matrix elements $P_S$ and $P_S^\dagger$ have a negligible effect on $T_1$ and $T_2$ and were therefore omitted. To separate the qubit subspace from $\ket{(0,2)S}$, we perform a fourth-order Schrieffer-Wolff transformation assuming that $|t|$ and $\tilde{P}$ are small compared to $U-\epsilon-V_+-|b_B|/2$. Then we apply a unitary transformation to the resulting 2$\times$2 Hamiltonian that diagonalizes the phonon-independent part, as it was done in Sec.~\ref{subsubsec:SimpleModel}. Consequently, the $\delta B_{\tilde{x}}$ and $\delta B_{\tilde{z}}$ that we derive from the qubit Hamiltonian characterize $T_1$ and $T_\varphi$, respectively [see Eqs.~\eqref{eq:T2}--\eqref{eq:Tphi}]. The parameters we use for our calculation allow us to assume $V_+\simeq V_-$ and simplify the expressions for $\delta B_{\tilde{x}}$ and $\delta B_{\tilde{z}}$ as follows:
\begin{eqnarray}
\label{eq:Bx2}
\delta B_{\tilde{x}}(\tau) &\simeq& \frac{b_B t^2 \left[b_B^2 t^2 + 2(t^2G_5^2+G_5^4-4t^4)\right]}{2 G_5 G_6}\tilde{P}(\tau) \nonumber \\  
& & \vspace{1cm} - \frac{b_B t^2 (G_5^2-t^2)}{G_6}\tilde{P}^2(\tau)  , \\
\label{eq:Bz2}
\delta B_{\tilde{z}}(\tau) &\simeq& \left[t^2G_5[4t^2(2t^2-G_5^2)+b_B^2(2G_5^2-3t^2)]\tilde{P}(\tau)\right. \nonumber \\  &&\left.+t^4[b_B^2+4(G_5^2-2t^2)]\tilde{P}^2(\tau)\right]\frac{1}{2G_5G_6},
\end{eqnarray}
where we introduced
\begin{eqnarray}
G_5&=&U-V_+-\epsilon,\\
G_6&=&G_5^3\sqrt{b_B^2(t^2-G_5^2)^2+\frac{t^4(b_B^2-8t^2+4G_5^2)^2}{4G_5^2}}.\ \ \ \ \ \
\end{eqnarray}
Using these expressions, we will now discuss the numerical data shown in Sec.~\ref{subsec:LargeDetuning}. 

As a first example, we start with the remarkable decay of $T_1$ by two orders of magnitude seen in Fig.~\ref{fig:dbb_dep}. In the dependence of $T_1$ on $b_B$ the two-phonon process is dominating, especially for smaller $b_B$ and larger temperature (for $b_B=2\mbox{ $\mu$eV}$ and $T=100\mbox{ mK}$, $\Gamma^{2p}_1/\Gamma^{1p}_1\simeq 2.7$). To analyze this dependence we therefore consider only two-phonon process terms in $\delta B_{\tilde{x}}$, i.e., the prefactor before $\tilde{P}^2(\tau)$. From Eq.~\eqref{eq:Bx2}, we see that the numerator of the prefactor is linear in $b_B$. The denominator is also a function of $b_B$, however it is of the form $\sqrt{C_{9}b_B^4 + C_{10} b_B^2 + C_{11}}$, where $C_{9}$, $C_{10}$, and $C_{11}$ are constants. Consequently, the power-law $T_1\propto b_B^{-2}$ holds very well for $b_B<1\mbox{ $\mu$eV}$ and slightly deviates for larger $b_B$. 

The dependence of $T_1$ on the detuning $\epsilon$ plotted in Fig.~\ref{fig:detuning_dep} is more complicated. Our numerical calculations show that in this case the two-phonon process again dominates. To understand the detuning-dependence of $T_1$ we therefore study the prefactor before $\tilde{P}^2(\tau)$ again. For the range of $\epsilon$ presented in Fig.~\ref{fig:detuning_dep}, the dependence of this prefactor on $\epsilon$ is approximately of the type $(U - V_{+} - \epsilon)^{-3}$, which suggests that $T_1\propto (U - V_{+} - \epsilon)^{6}$. As expected from this simple estimate, the relaxation time $T_1$ decreases rapidly with increasing $\epsilon$ in Fig.~\ref{fig:detuning_dep}. We recall that this estimate is solely based on the prefactor of $\tilde{P}^2(\tau)$ in Eq.~\eqref{eq:Bx2}. Corrections to the detuning dependence of $T_1$ can, e.g., be expected from the factor $\cos(\tau \Delta_{ST} / \hbar)$ in the integral of Eq.~\eqref{eq:Jplus}. As mentioned in Sec.~\ref{subsubsec:ST0LargeDetDepOnDet}, the splitting $\Delta_{ST}$ is strongly affected by $\epsilon$ for the parameters of Fig.~\ref{fig:detuning_dep}.

The dependence on the tunnel coupling $t$ is very complex. As we see from Figs.~\ref{fig:t_dep} and \ref{fig:rates_t_dep}, both one- and two-phonon processes contribute significantly to $T_1$ and $T_2$. However, Eqs.~\eqref{eq:Bx2} and \eqref{eq:Bz2} can be greatly simplified when focusing on certain regimes. For example, we see from Fig.~\ref{fig:rates_t_dep} that the two-phonon process dominates in $T_1$ for $T=500\mbox{ mK}$. When we analyze the prefactor before $\tilde{P}^2(\tau)$ in Eq.~\eqref{eq:Bx2}, we find that its dependence on $t$ is relatively weak for $3\mbox{ $\mu$eV} \leq t \leq 8\mbox{ $\mu$eV}$, which is consistent with $\Gamma_1^{2p}$ in Fig.~\ref{fig:rates_t_dep}.

\subsection{Zero detuning}
\label{subsec:SmallDetuning}

For the case of zero detuning, i.e., $\epsilon\simeq 0$, we have to take into account the first excited orbital states. We will therefore consider our Hamiltonian in the basis $\{\ket{(1,1)S}$, $\ket{(1,1)T_0}$, $\ket{(1,1)T_+}$, $\ket{(1,1)T_-}$, $\ket{(1^*,1)S}$, $\ket{(1^*,1)T_+}$, $\ket{(1^*,1)T_0}$, $\ket{(1^*,1)T_-}\}$, where the asterisk indicates that the electron in the QD is in the first excited state \cite{kornich:prb14},
\begin{widetext}
\begin{equation}
\widetilde{H}=
\begin{pmatrix}
    -J_S + P_{SS} & \frac{b_B}{2} & \frac{\Omega}{\sqrt{2}} & -\frac{\Omega}{\sqrt{2}} & P^e_{cr} & \frac{\Omega_1}{\sqrt{2}} & 0 & -\frac{\Omega_1}{\sqrt{2}} \\
    \frac{b_B}{2} & P_T & 0 & 0 & 0 & -\frac{\Omega_1}{\sqrt{2}} & P^e_{cr} & -\frac{\Omega_1}{\sqrt{2}}  \\ 
    \frac{\Omega}{\sqrt{2}} & 0 & E_Z + P_T & 0 & \frac{\Omega_1}{\sqrt{2}} & P^e_{cr} & - \frac{\Omega_1}{\sqrt{2}} & 0 \\
    -\frac{\Omega}{\sqrt{2}} & 0 & 0 & -E_Z+P_T & -\frac{\Omega_1}{\sqrt{2}} & 0 & -\frac{\Omega_1}{\sqrt{2}} & P^e_{cr}\\
    P^{e\dag}_{cr} & 0 & \frac{\Omega_1}{\sqrt{2}} & -\frac{\Omega_1}{\sqrt{2}} & \Delta E +P^e & \frac{\Omega_2}{\sqrt{2}} & 0 & -\frac{\Omega_2}{\sqrt{2}}\\
    \frac{\Omega_1}{\sqrt{2}} & -\frac{\Omega_1}{\sqrt{2}} & P_{cr}^{e\dag} & 0 & \frac{\Omega_2}{\sqrt{2}} & \Delta E+E_Z+P^e & -\frac{\Omega_3}{\sqrt{2}} & 0\\
    0 & P^{e\dag}_{cr} &  - \frac{\Omega_1}{\sqrt{2}} & -\frac{\Omega_1}{\sqrt{2}} & 0 & -\frac{\Omega_3}{\sqrt{2}} & \Delta E +P^e & -\frac{\Omega_3}{\sqrt{2}} \\
    -\frac{\Omega_1}{\sqrt{2}} & -\frac{\Omega_1}{\sqrt{2}} & 0 &  P^{e\dag}_{cr} & -\frac{\Omega_2}{\sqrt{2}} & 0 & -\frac{\Omega_3}{\sqrt{2}} & \Delta E-E_Z+P^e\\
  \end{pmatrix} + H_{ph} .
\label{eq:Matrix8x8zeroDetuning}
\end{equation}
\end{widetext}
Here, the splitting $J_S$ takes into account the hybridization of $\ket{(0,2)S}$ and $\ket{(2,0)S}$ with $\ket{(1,1)S}$ and is defined as
\begin{equation}
J_S=\frac{1}{2}(\sqrt{16 t^2+(U-V_+)^2}-U-V_++2V_-).
\end{equation}
The matrix elements $P^e$, $P^e_{cr}$, $P^{e\dagger}_{cr}$ result from the electron-phonon interaction in the same way as was shown in Sec.~\ref{subsec:BasisStates}, but for the corresponding excited states. The matrix element $P_{SS}$ is a linear combination of electron-phonon interaction matrix elements including the effect of $\ket{(2,0)S}$ and $\ket{(0,2)S}$. The terms $\Omega_1, \Omega_2, \Omega_3$ arise from SOI. The derivation of all these matrix elements is described in detail in Ref.~\onlinecite{kornich:prb14}, Appendix~C.

\begin{figure}[tb] 
\begin{center}
\includegraphics[width=\linewidth]{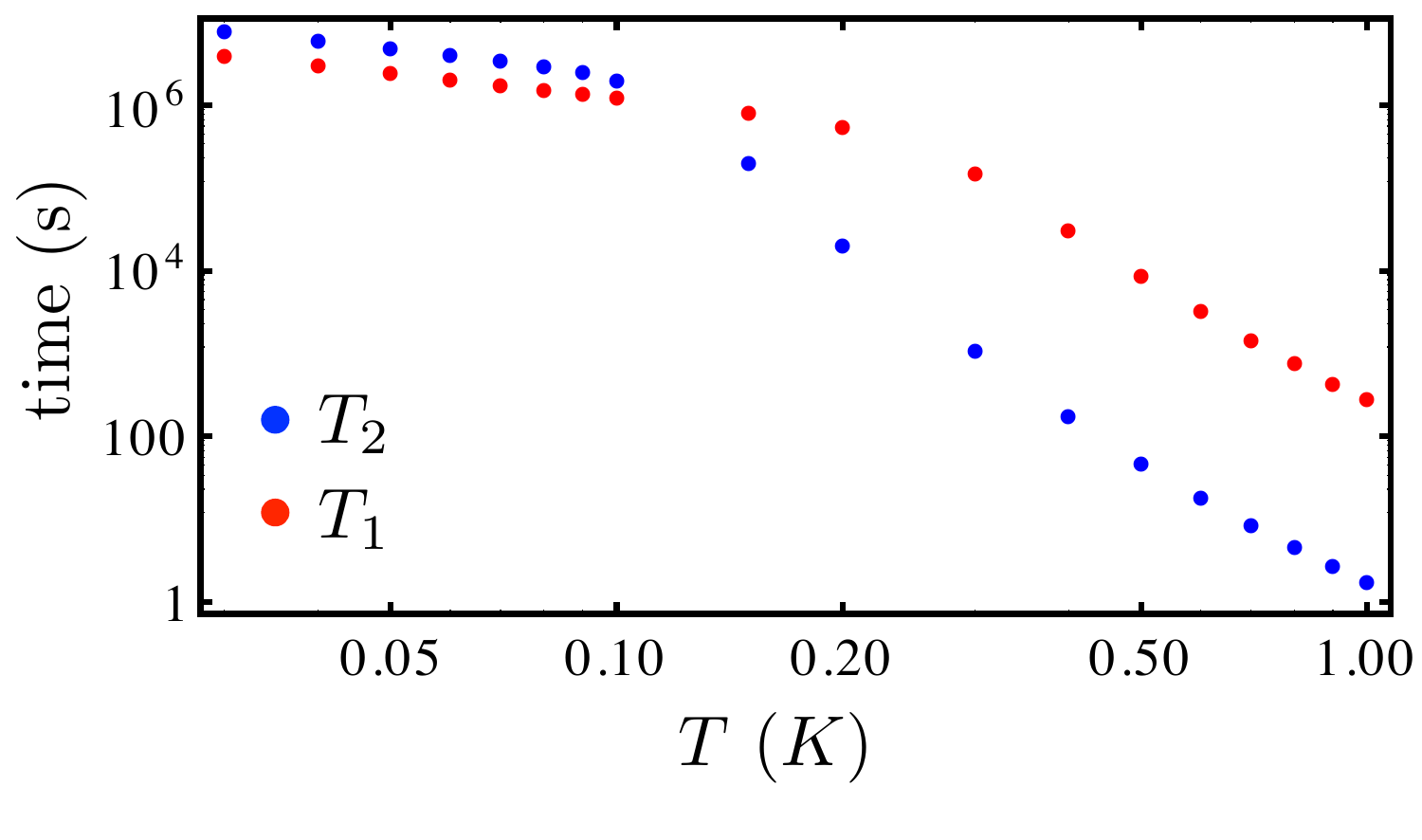}
\caption{The dependence of $T_2$ and $T_1$ of a $S$-$T_0$ qubit on temperature for the unbiased case $\epsilon \simeq 0$. The parameters are provided in Sec.~\ref{subsec:SmallDetuning}. }
\label{fig:temp_dep_zero_det}
\end{center}
\end{figure}

We then perform an initial unitary transformation, followed by a Schrieffer-Wolff transformation, and apply Bloch-Redfield theory as described in Sec.~\ref{subsec:BlochRedfieldTheory} and plot the temperature dependence of $T_1$ and $T_2$, see Fig.~\ref{fig:temp_dep_zero_det}. Here we take $B=0.4\mbox{ T}$, $t=24\mbox{ $\mu$eV}$, $U=1.2\mbox{ meV}$, $V_+=50\mbox{ $\mu$eV}$, $V_-=49.5\mbox{ $\mu$eV}$, $\Delta E=200\mbox{ $\mu$eV}$, $L=150\mbox{ nm}$, $l_R=2\mbox{ $\mu$m}$, and $b_B=-1\mbox{ $\mu$eV}$. Consequently, $J_{S}=1.5\mbox{ $\mu$eV}$ and $\Delta_{ST}=2.5\mbox{ $\mu$eV}$. Comparing Fig.~\ref{fig:temp_dep_zero_det} with Fig.~\ref{fig:temp_dep_large_det} we see that the qubit lifetimes are several orders of magnitude longer than in the case of large detuning. This makes the zero detuning regime favorable for $S$-$T_0$ qubits, which was also shown for DQDs in GaAs/AlGaAs in our previous work \cite{kornich:prb14}.

The calculations for Fig.~\ref{fig:temp_dep_zero_det} were done with the orbital excitation along the axis that connects the QDs. The decay rates resulting from excitation along the orthogonal direction do not change the qualitative picture, which is sufficient for our consideration. For Fig.~\ref{fig:temp_dep_zero_det} we chose $\eta=0$. If we take $\eta=\pi/2$, the rates are either smaller or of the same order as for $\eta=0$. States of type $(1,1^*)$ with the excited electron in the right QD will change the results only by factors around 2, and therefore were not included for simplicity. 

The valley degrees of freedom were neglected in our model because valley splittings around 1~meV were already realized experimentally \cite{goswami:nature07, xiao:apl10, yang:natcom13}, which is a large gap compared to the orbital level spacing $\Delta E=200\mbox{ $\mu$eV}$. While valley-related effects are strongly suppressed when the valley splitting is large, we note that they can be a significant source of decoherence when the splitting is small \cite{tahan:prb02, culcer:prb10, tahan:prb14, gamble:prb12, rohling:njp12}. Therefore, setups with a large valley splitting are usually favorable when implementing spin qubits in Si/SiGe heterostructures, which is the case that we focus on in this work.

\section{Comparison with other decay mechanisms}
\label{sec:Comparison}

In our previous calculations for $S$-$T_0$ qubits in GaAs DQDs \cite{kornich:prb14}, we found that the considered one- and two-phonon processes may very well correspond to the dominant decay channels in an experiment. In contrast, for the Si DQDs studied here, the obtained decay times for singlet-triplet qubits are relatively long, at least for many parameter regimes. We note that this finding is consistent with a recent analysis of resonant exchange qubits in three-electron triple quantum dots \cite{srinivasa:prb16}, where the relaxation times due to phonons were predicted to be orders of magnitude longer in Si than in GaAs. Consequently, it is well possible that the experimentally feasible qubit lifetimes will be limited by other mechanisms, some of which we briefly discuss below. Nevertheless, even if other mechanisms turn out to dominate in standard regimes for qubit operation, we identified and proposed ways how our theory can be confirmed experimentally, which would be a desirable contribution to understanding and assessing the role of the discussed one- and two-phonon processes in Si-based systems.

Among the most relevant noise sources for electrically controllable qubits is charge noise \cite{kuhlmann:nphys13, dial:prl13, kim:nnano15}, which may be due to charge traps within the heterostructure or noise from the gates. For instance, electrical noise was considered as a major obstacle for the implementation of high-quality two-qubit gates between $S$-$T_0$ qubits in GaAs \cite{shulman:science12}. Theoretical studies suggest that the effects of charge noise in GaAs and Si are similar to a great extent \cite{hu:prl06, culcer:apl09}. As evident from, e.g., the pure-dephasing model discussed in Ref.~\onlinecite{cywinski:prb08}, the resulting decay will depend both on the spectral density of the noisy fluctuations in the level splitting of the qubit and on the details of the operation scheme, as suitable pulse sequences for dynamical decoupling may strongly prolong the dephasing time \cite{viola:prl99, petta:science05, barthel:prl10, bluhm:nature11, shulman:science12, kim:npj15}. Furthermore, decoherence due to charge noise can be much suppressed by operating the qubit at a sweet spot, where the level splitting of the qubit is insensitive to electric field fluctuations. This is a particularly advantageous feature of $S$-$T_0$ qubits in unbiased DQDs \cite{loss:pra98, burkard:prb99, reed:arxiv15, martins:arxiv15} and $S$-$T_-$ qubits (especially those based on $\ket{(1,1)T_-'}$-$\ket{(1,1)S'}$, see Sec.~\ref{subsec:Tminus11S}) operated at the anticrossing \cite{chesi:prb14, wong:prb15}.

While we considered here the Bloch-Redfield theory and studied the phonon-assisted relaxation and decoherence that results from one- and two-phonon processes, such as the two-phonon Raman process \cite{mccumber:jap63, yen:pr64, altner:prb96, meltzer:book05, roszak:prb09, kornich:prb14er}, a spin-boson model was adopted in Ref.~\onlinecite{hu:prb11} in order to describe pure dephasing of $S$-$T_0$ qubits in the absence of any real or virtual phonon absorption or emission. In the calculations of Ref.~\onlinecite{hu:prb11}, interactions between the electrons and a dissipative phonon reservoir lead to an exponential decay of the qubit coherence, and the associated dephasing time depends strongly on the overlap of the electron wave functions and the decay properties of the phonon bath. In contrast to our model, where the qubit lifetimes in GaAs turned out to be limited by the piezoelectric electron-phonon coupling \cite{kornich:prb14}, the lifetimes calculated in Ref.~\onlinecite{hu:prb11} for both Si and GaAs are limited by the deformation potential coupling. Depending on the experimental setup this additional decay channel might dominate, particularly for strongly overlapping quantum dots in Si, and it can be suppressed by moving the two dots farther apart \cite{hu:prb11}.

As mentioned before in Sec.~\ref{subsec:SmallDetuning}, valley-related effects can become an important source of decoherence if the energy splitting between valleys is not sufficiently large \cite{tahan:prb02, culcer:prb10, tahan:prb14}. Among other things, disorder or interface effects for Si/SiGe and Si/SiO$_2$ can play a significant role here \cite{saraiva:prb11, culcer:prb10, nestoklon:prb06, chutia:prb08, gamble:prb13, tahan:prb14}. When the valley splitting is large, however, qubit decoherence due to the valley degrees of freedom is suppressed, and splittings of the order of 1~meV or even more are experimentally feasible \cite{goswami:nature07, xiao:apl10, yang:natcom13, zwanenburg:rmp13, boykin:apl04}.

Finally, the coherence of qubits in Si/SiGe heterostructures can be lost due to interaction with the nuclear spins, although the hyperfine-induced dephasing time of 360\mbox{ ns} (no echo pulses) reported for a Si DQD \cite{maune:nature12} is already one to two orders of magnitude longer than the typical values for GaAs \cite{khaetskii:prl02, merkulov:prb02, coish:prb04, petta:science05}. Ultimately, however, the hyperfine coupling will not present a limiting factor for the qubit lifetimes, since Si and Ge can be grown nuclear-spin-free.

\section{Conclusions}
\label{sec:Conclusions}

We considered $S$-$T_-$ qubits in the anticrossing region for the two cases where the singlet is mainly $\ket{(0,2)S}$ and where it is mainly $\ket{(1,1)S}$. In the latter case, $T_1$ and $T_2$ turned out to be much longer than in the former one. We showed that the magnetic field gradient reduces $T_1$ and $T_2$ substantially, when it is above a certain value at which the one-phonon process starts to dominate over the two-phonon process. This follows from the fact that the magnetic field gradient provides the splitting in the anticrossing, and therefore the one-phonon process is very sensitive to its change. In contrast, two-phonon-based relaxation does not change noticeably in the range of parameters we use, and two-phonon-based dephasing is very weak even though it does depend on the magnetic field gradient. We proposed regimes where our theory of one- and two-phonon processes may be experimentally tested. Remarkably, $T_2$ ($T_1$) has a peak (dip)
at the center of the $S$-$T_-$ anticrossing in the dependence on the applied magnetic field (Fig.~\ref{fig:B_dep_twoph}). As the external magnetic field can easily be changed in an experiment, this peak (dip) might be an experimental indication of the center of the anticrossing, which is a regime of interest e.g. for Refs.~\onlinecite{chesi:prb14, wong:prb15}.

We also studied $S$-$T_0$ qubits in the regimes which were presented in our previous work on DQDs in GaAs/AlGaAs \cite{kornich:prb14}, i.e., at large detuning in the anticrossing region of the singlets and at zero detuning. The key result that small detuning is much more favorable regarding the qubit lifetimes than large detuning is valid here too. We showed that in the anticrossing region even small changes in $\epsilon$ may shorten $T_1$ and $T_2$ by two orders of magnitude. We note that the relation $T_\varphi\ll T_1$, shown in our previous work for the regime of large detuning, does not hold for the usual parameters of experiments with SiGe/Si/SiGe DQDs because of a rather large applied magnetic field gradient. 
We showed that the magnetic field gradient can reduce $T_1$ by orders of magnitude. We demonstrated that the dependence of $T_1$ on tunnel coupling is qualitatively different for different temperatures, which is explained by the behavior of one- and two-phonon processes. 
Our study of the effect of various system parameters on $T_1$ and $T_2$ shows ways how to prolong the phonon-based decoherence and relaxation times by orders of magnitude.

\begin{acknowledgments} 
We thank S.\ Chesi, S.\ N.\ Coppersmith, M.\ Friesen, T.\ Otsuka, K.\ Takeda, and J.\ R.\ Wootton for helpful discussions and acknowledge support from the Swiss National Science Foundation, NCCR QSIT, SiSPIN, the ITNs S$^3$NANO and Spin-NANO, and IARPA.  
\end{acknowledgments}

\bibliographystyle{plain}

\end{document}